\documentclass[10pt,twocolumn,twoside]{IEEEtran}

\usepackage{graphicx,amssymb,amsmath,amsbsy,bm,amsfonts,hyperref}

\usepackage{pdfpages}

\graphicspath{{figures/}}

\newtheorem{remark}{Remark}
\newtheorem{definition}{Definition}

\newcommand{\lrw}{L_{rw}}
\newcommand{\lsym}{L_{sym}}
\newcommand{\dmax}{d_{\max}}
\newcommand{\dmin}{d_{\min}}
\newcommand{\C}[2]{\mathcal{C}_{#1, #2}}

\newcommand\numberthis{\addtocounter{equation}{1}\tag{\theequation}}

\newcommand{\eqnref}[1]{(\ref{#1})}

\def\EqualDef{\mathrel{\mathop=^{\rm def}}}
\def\lsupp{\ell}

\begin{document}

\title{Comparing Graph Spectra of Adjacency and Laplacian Matrices}

\author{J.~F.~Lutzeyer \&  A.~T.~Walden
\date{\today}
\thanks{
Copyright (c) 2017 IEEE. Personal use of this material is permitted. However, permission to use this material for
any other purposes must be obtained from the IEEE by sending a request to pubs-permissions@ieee.org. 
Johannes Lutzeyer and Andrew Walden
are both at the Department of Mathematics, Imperial College  London, 180 Queen's Gate,
London SW7 2AZ, UK.  
(e-mail:  johannes.lutzeyer11@imperial.ac.uk and a.walden@imperial.ac.uk).} }
\IEEEpubid{}
\maketitle
\begin{abstract}
Typically, graph structures are represented by one of three different matrices: the adjacency matrix, the unnormalised and the normalised graph Laplacian matrices. The spectral (eigenvalue) properties of these different matrices are compared. For each pair, the comparison is made by applying an affine transformation to one of them, which enables comparison whilst preserving certain key properties such as normalised eigengaps. Bounds are given on the eigenvalue differences thus found, which depend on the minimum and maximum degree of the graph. The monotonicity of the bounds and the structure of the graphs are related. The bounds on a real social network graph, and on three model graphs, are illustrated and analysed. The methodology is extended to provide bounds on normalised eigengap differences which again turn out to be in terms of the graph's degree extremes. 
It is found that if the degree extreme difference is large, different choices of representation matrix may give rise to disparate inference drawn from graph signal processing algorithms; smaller degree extreme differences result in consistent inference, whatever the choice of representation matrix. The different inference drawn from signal processing algorithms is visualised using the spectral clustering algorithm on the three representation matrices corresponding to a model graph and a real social network graph.

\end{abstract}
\section{Introduction}

Graph structures are usually represented by one of three different matrices: the adjacency matrix, and unnormalised and normalised graph Laplacian matrices. We will refer to these three matrices as representation matrices. In recent years there has been an increasing interest in the use of graph structures for modelling purposes and their analysis.
An overview of the numerous advances in the area of signal processing on graphs is provided in \cite{shuman2013}. 
\cite{cvetkovic2011} contains a vast bibliography and overview of the applications of graph spectra in engineering, computer science, mathematics and other areas.
\cite{sandryhaila2014} address the big data topic in the context of graph structures by introducing different product graphs and showing how discrete signal processing is sped up by utilising the product graph structure.

In a particular analysis, given a choice of one representation matrix, the  spectral (eigenvalue) properties of the matrix are often utilised as in, for example, the spectral clustering algorithm \cite{vonLuxburg2007} and graph wavelets \cite{tremblay2014}. 
Spectral clustering has had a big impact in the recent analysis of graphs,
with active research on the theoretical guarantees which can be given for a clustering determined in this way (\cite{lei2015},\cite{rohe2011}). \cite{tremblay2014} use their construction of graph wavelets, utilising both the spectrum and the eigenvectors in their design, to obtain clusters on several different scales. 

The different representation matrices are well established in the literature.
A shifted version of the symmetric normalised graph Laplacian is used in the analysis by \cite{rohe2011}. \cite{shuman2013} work with the unnormalised graph Laplacian $L$ and remark on both normalised graph Laplacians $\lrw$ and $\lsym,$ while \cite{sandryhaila2014} use the adjacency matrix $A$ and remark on the unnormalised graph Laplacian $L$. 
\cite{tremblay2014} use the symmetric normalised graph Laplacian $\lsym$ in the construction of their graph wavelets. \cite{lei2015} perform their analysis using the adjacency matrix $A$ and mention the extension of their results to the graph Laplacian as future work. The question of what would have happened if another representation had been chosen instead, arises naturally when reading the networks and graph literature. The purpose of this paper is to compare the spectral  properties of these different matrices, to try to gain some insight into this issue.

Our approach is to firstly apply an affine transformation to one of the matrices, before making the comparison. The transformation preserves certain key properties,  such as normalized eigengaps (differences between successive eigenvalues) and the ordering of eigenvalues.

Consider a graph $G = (V,E)$ with vertex set $V$, where $\lvert V \rvert =n $, and edge set $E$.  We will assume our graphs to be undirected and simple. An edge between vertices $v_i$ and $v_j$ has edge weight $w_{ij}$, which is defined to be 0 if there exists no edge between $v_i$ and $v_j$. If the edge weights are binary, then the graph is said to be unweighted, otherwise it is a weighted graph. The matrix holding the edge weights is  the adjacency matrix $A \in \left[0,1 \right]^{n \times n}$ of a graph and is defined to have entries $A_{ij}=w_{ij},$ (the maximal weight of an edge, unity,  can be achieved by normalising by the maximal edge weight if necessary). The adjacency matrix is one of the standard graph representation matrices considered here.

The degree $d_i$ of the $i^\mathrm{th}$ vertex is defined to be the sum of the weights of all edges connecting this vertex with others, i.e., $d_i = \sum_{j=1}^n w_{ij}$ and the degree matrix $D$ is defined to be the diagonal matrix $D = \mathrm{diag}(d_1,\ldots,d_n)$. The degree sequence is the set of vertex degrees $\{d_1,\ldots, d_n\}$. The minimal and maximal degree in the degree sequence is denoted $\dmin$ and $\dmax$, respectively. (An important special case is that of $d$-regular graphs, in which each vertex is of degree $d,$
so that  $d_{\text{max}} = d_{\text{min}} =d,$ and  the degree matrix is a scalar multiple of the identity matrix, $D= d I$,
 where $I$ is the identity matrix.)

For a general $D$ matrix, the unnormalised graph Laplacian, $L,$ is defined as 
$$L=D-A.$$
 In the literature there are two normalised Laplacians which are considered \cite{vonLuxburg2007}; these follow from the unnormalised graph Laplacian by use of the inverse of the degree matrix as follows: 
\begin{equation}\label{eq:nLaplacians}
L_{rw} = D^{-1} L; \qquad L_{sym} = D^{-1/2} L D^{-1/2}.
\end{equation}
Since they are related by a similarity transform 
the sets of eigenvalues of these two normalised Laplacians are identical 
and we only need to concern ourselves with one: we work with $L_{rw}.$ 

The three representation matrices we utilise in this paper are thus
$A, L $ and $L_{rw}.$

We use the common nomenclature that the set of eigenvalues of a matrix is called its spectrum. In order to distinguish eigenvalues of $A, L$ and $\lrw$ we use $\mu$ for an eigenvalue of $A$, $\lambda$ for an eigenvalue of $L$ and $\eta$ for one of $\lrw.$ 
We write $\lambda({\hat L})$ to denote an eigenvalue of a perturbation ${\hat L}$ of $L,$
and likewise for the other matrices.

\subsection{Motivation of the affine transformations}

For $d$-regular graphs, since $D= d I$, the spectra of the three graph representation matrices are exactly related via known affine transformations, see for example \cite[p.~71]{vanMieghem2011}. For general graphs, the relation of the representation spectra is non-linear. The transformation between representation spectra can be shown to be either non-existent, if repeated eigenvalues in one spectrum correspond to unequal eigenvalues in another spectrum, or to be polynomial of degree smaller or equal than $n$, see Appendix \ref{app_poly_relation}. In practice, finding the polynomial mapping between spectra is subject to significant numerical issues, which are also outlined in Appendix \ref{app_poly_relation}; furthermore, crucial spectral properties such as eigengap size and eigenvalue ordering are not preserved by these non-linear transformations. Importantly, in the spectral clustering algorithm, eigengaps are used to determine the number of clusters present in the network, while the eigenvalue ordering, an even more fundamental property, 
is used to identify which eigenvectors to consider \cite{vonLuxburg2007}. Therefore, when comparing the impact of the choice of the representation matrices on graphical analysis, we need to preserve eigenvalue ordering and eigengap size. 

In this paper we consider affine transformations between spectra corresponding to general graphs and bound the resulting eigenvalue differences. Affine transformations preserve eigenvalue ordering and eigengap size. Further details on the properties of the affine transformations are given in Section \ref{sec_tranf_properties}.

Comparing the representation spectra without transforming first or only using an additive constant will not yield an informative comparison. This is due to the ordering of eigenvalues being reversed between the adjacency matrix and the Laplacian matrices, i.e., the largest eigenvalue of the adjacency matrix corresponds to the smallest eigenvalue of the Laplacians (see Section \ref{sec_bd_A_L}). Therefore, a reflection of one of the spectra using a multiplicative constant is a sensible preprocessing step before comparing spectra. Furthermore, using an affine transformation has the advantage that we are able to map the spectral supports of the three representation matrices onto each other to further increase comparability of the spectra. This mapping of spectral supports requires scaling.

Let $X$ and $Y$ be any possible pair of the three representation matrices to be compared. Our general approach is to define an affine transform of one of them, say ${\cal F}(X).$  We then study bounds on the eigenvalue error 
\begin{equation*}
|\lambda_i({\cal F}(X))-\lambda_i(Y)|,
\end{equation*}
where $\lambda_i(Y)$ is the eigenvalue of $Y.$

\begin{remark}\label{remark:dreg}
For $d$-regular graphs our transformations recover the exact relations between the spectra of the three matrices. This is  a crucial baseline to test when considering maps between the representation spectra.\hfill$\square$
\end{remark}

\subsection{Contributions}
The contributions of this paper are:
\begin{enumerate}
\item A framework under which the representation matrices can be compared.  
Central to this framework is an affine transformation, mapping one matrix as closely as possible on to the other in spectral norm. This procedure is rather general and is likely to be applicable in other matrix comparisons. Properties of these transforms, such as normalised eigengap preservation, are discussed.
\item The quantification of the spectral difference between the representation matrices at an individual eigenvalue level, and at eigengap level, via closed form bounds.
\item A partition of graphs, according to their degree extremes, derived from the bounds, which enables an interesting analysis of our results.
\item The recognition that if the degree extreme difference is large, different choices of representation matrix may give rise to disparate inference drawn from graph signal processing algorithms; smaller degree extreme differences will result in consistent inference, whatever the choice of representation matrix. 
\end{enumerate}

In Sections~\ref{sec_bd_A_L}, \ref{sec_bd_L_Lrw} and \ref{sec_bd_A_Lrw},  bounds are provided for the eigenvalue difference between any two of the three representation matrices, $A$, $L$ and $\lrw$. 
The properties of the transformation used in each comparison are elaborated in Section \ref{sec_tranf_properties}. The analysis of the bound values over all graphs partitioned by their degree extremes is given in Sections \ref{sec_relating_bounds} and \ref{sec_explaining_structure_of_tab}. Section \ref{sec_visualising_data} displays a proof of concept by applying the bounds to graphs arising from a social network and three further model examples, and finds tightness of one of the bounds on two of the examples. In Section~\ref{sec_eigengap_bound} the bound on normalised eigengap differences for each of the pairs of representation matrices is derived. These bounds are illustrated via examples with varying degree extreme difference in Section \ref{sec_eigengap_bound_examples}. In Section \ref{sec_clustering} we present the spectral clustering of several graphs to illustrate the effects of the representation matrix choice in practice. Section~\ref{sec:summary} provides a summary and conclusions.

\section{Bounding the individual eigenvalue difference for $A$ and $L$} \label{sec_bd_A_L}

Let $A$ have eigenvalues $\mu_n\leq \cdots \leq \mu_1.$  Then, with $\iff$ denoting `if and only if,' for any eigenvalue $\mu$ and  eigenvector $w,$ we have, for $d_1\in\mathbb{R},$
\begin{align*}
A w &= \mu w&\\
\iff \left( d_1 I - D + D -A \right) w &= \left( d_1 - \mu \right) w &\\
\iff \left( d_1 I - D + L \right) w &= \left( d_1 - \mu \right) w\\
\iff \hat{L} w &= \lambda(\hat{L}) w,
\end{align*}
where
\begin{eqnarray}
\hat{L} &{\displaystyle{\EqualDef}}&  d_1 I-D +L=d_1 I-A \label{eqn_defn_L_hat}\\
\lambda(\hat{L}) &{\displaystyle{\EqualDef}}& d_1- \mu.  \label{eq1_Linverted}
\end{eqnarray}
$\hat{L}$ is an affine transform ${\cal F}(A)$ of $A$ with transform parameter $d_1.$ It can also be viewed as a perturbation of $L$ with diagonal `error' $d_1I-D,$
allowing us to find an error bound on the accuracy of the eigenvalue approximation of the unnormalised Laplacian eigenvalues, $\lambda_i,$ by the eigenvalues of $\hat{L},$ denoted $\lambda_i(\hat{L})$. In general $\hat{L}$ does not itself have graph Laplacian structure, since, for non-$d$-regular graphs, not all row sums of $\hat{L}$ are equal to zero, a property of a graph Laplacian.
A multiplicative free parameter, in addition to the parameter $d_1,$ yields no improvement of the final result, and is hence omitted.

In order to relate the eigenvalues of the adjacency matrix $A$ and unnormalised graph Laplacian $L,$ we look at the error in the direct spectral relation via $\hat{L}$. From (\ref{eq1_Linverted}), we can see that the spectrum in this direct relation gets reordered, i.e., the larger end of the spectrum of the adjacency matrix $A$ corresponds to the smaller end of $\hat{L}$.  Therefore, we will order the eigenvalues of the unnormalised graph Laplacian in opposite order to the ones of the adjacency matrix and denote them by $\lambda_1  \leq \cdots \leq \lambda_n,$ with corresponding eigenvectors $v_1, \ldots, v_n$.

In order to obtain a bound on the eigenvalue error $\left\lvert \lambda_i( \hat{L}) - \lambda_i \right\rvert,$
we will utilise the spectral norm $\left\| \cdot \right\|_2.$    Since $A$, $L$ and $\lsym$ are all real and symmetric, we pay particular attention to real, symmetric matrices, noting again that $\lsym$ and $\lrw$ have the same eigenvalues.

\begin{definition}
The spectral norm, or Euclidean 2-norm, of a matrix $B,$ is defined to be the largest singular value of $B$, i.e.
$
\left\| B \right\|_2 \EqualDef [{\rm{max~eigenvalue~of~}}\left(B^T B\right)]^{1/2}.
$
For any real, symmetric matrix, singular values and eigenvalues coincide in absolute value.  
Hence, we can express the spectral norm of a real, symmetric matrix $C,$ with ordered eigenvalues 
$\lambda_1(C) ,\ldots,\lambda_n(C),$ as
$
\left\| C \right\|_2 = \max\left( \lvert \lambda_1(C) \rvert, \lvert \lambda_n(C) \rvert \right).
$ 
\hfill$\square$
\end{definition}

Now, we proceed to analyse the error made by approximating $\lambda_i$ by $\lambda_i(\hat{L})$. Firstly, \cite[eqn. (4.3), Chapter~4.1]{bai2000}
\begin{equation}
\left\| \hat{L} - L \right\|_2 \geq \left\lvert \lambda_i( \hat{L}) - \lambda_i \right\rvert =|d_1-\mu_i-\lambda_i|,  \label{eqn_eval_difference}
\end{equation}
for $ i \in \left\{ 1,2, \ldots, n\right\},$
i.e., the difference of the eigenvalues of a symmetric matrix $L$  and its perturbed version $\hat{L}$ is bounded by the two norm of the perturbation $\hat{L} - L $. Now,
\begin{eqnarray}
\left\| \hat{L} - L \right\|_2 &\overset{\eqref{eqn_defn_L_hat}}{=}& \left\| d_1I - D \right\|_2\nonumber\\
&=& \text{max}\left( \left\lvert d_1-d_{\text{max}}  \right\rvert, \left\lvert d_1-d_{\text{min}}  \right\rvert \right) \label{eqn_defn_d}\\
&=& \frac{d_{\text{max}} - d_{\text{min}}}{2}\nonumber,
\end{eqnarray}
where $ d_1 = ({d_{\max} + d_{\min}})/{2} .$ (\ref{eqn_defn_d}) uses the fact that the eigenvalues of diagonal matrices are equal to their diagonal elements. Furthermore, $d_1$ is chosen to minimise the upper bound on the approximation error. 
So, our affine transformation and parameter choice can be conceptualized as
$$
A \rightarrow {\cal F}_1(A) \EqualDef\hat{L} \approx L,
$$
where ${\cal F}_1(A)$ is the affine transform specified in (\ref{eqn_defn_L_hat}). 
Putting (\ref{eqn_eval_difference}) and 
(\ref{eqn_defn_d}) together,
\begin{equation}
\left\lvert \lambda_i( \hat{L}) - \lambda_i \right\rvert \leq \frac{d_{\text{max}} - d_{\text{min}}}{2} \EqualDef e(A,L),
\label{eqn_L_error_bound}
\end{equation}
where  we denote the final (error) bound  by $e(A,L)$.

\begin{remark} \label{rmk_L_weyl}
The result in (\ref{eqn_L_error_bound}) can also be obtained from Weyl's inequality, as shown in Appendix~\ref{app_weyl_bound}.\hfill$\square$
\end{remark}

For $d$-regular graphs, 
 (\ref{eqn_L_error_bound}) gives $ e(A,L)=0,$ so that 
$\lambda_i( \hat{L}) = \lambda_i$ and  (\ref{eq1_Linverted}) then gives
$\lambda_i=d_1-\mu_i=d-\mu_i,$
i.e., the eigenvalues are related by the required exact relation, as claimed in Remark~\ref{remark:dreg}.

For general graphs, using (\ref{eqn_L_error_bound}),  we can establish a rough correspondence, 
to within an affine transformation,
between the eigenvalues of the adjacency matrix, $A,$ and the unnormalised graph Laplacian, $L,$ if the extremes of the degree sequence $d_{\text{max}}$ and $d_{\text{min}}$ are reasonably close.

\section{Bounding the individual eigenvalue difference for $L$ and $L_{rw}$}\label{sec_bd_L_Lrw}
We now look at the spectral relationship between $L$ and  $\lrw$. 
As in Section~\ref{sec_bd_A_L}, let $L$ have eigenvalues $\lambda$ and eigenvectors $v.$ 
Then, with $c_1\in \mathbb{R},$
\begin{eqnarray*}
L v &=& \lambda v\\
\iff c_1 D D^{-1} L v &=& c_1 \lambda v  \\
\iff c_1 D L_{rw} v &=& c_1 \lambda  v \\
\iff \hat{L}_{rw} v &=& \eta(\hat{L}_{rw}) v, 
\end{eqnarray*}
where
\begin{eqnarray}
\hat{L}_{rw} &{\displaystyle{\EqualDef}}& c_1 D L_{rw} =c_1 L\label{eqn_defn_hat_lrw}\\ 
\eta(\hat{L}_{rw}) &{\displaystyle{\EqualDef}}& c_1 \lambda\label{eqn_secondmap}.
\end{eqnarray}

$\hat{L}_{rw}$ is an affine transform ${\cal F}(L)$ of $L$ with transform parameter $c_1.$ It can also be viewed as a multiplicative perturbation of $\lrw,$ via perturbation factor $c_1 D.$
Therefore, we also choose a scaling as the eigenvalue transformation and have $c_1$ as the only free parameter. (Adding a free translation can easily be shown to yield no improvement of the final bound.) The multiplicative transformation between matrices also rules out Weyl's inequality for finding the bound (see Remark~\ref{rmk_L_weyl}).

The following result is the equivalent of (\ref{eqn_eval_difference}):
\begin{equation}\label{eq:equiv}
\left\| \hat{L}_{rw} - L_{rw}\right\|_2  \geq \left\lvert \eta_i(\hat{L}_{rw}) - \eta_i\ \right\rvert = \left|c_1\lambda_i-\eta_i\right|,
\end{equation}
for $i \in \{1,2,\ldots,n\}.$
In the steps that follow we use firstly  that the eigenvalues of $\lrw$ are all positive, and secondly that  $\left\| L_{rw}\right\|_2 \leq 2;$ as established in \cite[p.~64]{vanMieghem2011}. Now,
\begin{eqnarray}
\!\!\!\!\!\!\!\!\left\| \hat{L}_{rw} - L_{rw}\right\|_2\!\!\!\!\!\! &=& \!\!\!\!\left\|  \left( c_1 D - I \right) L_{rw} \right\|_2\nonumber\\
\!\!\!\!\!\! &\leq& \!\!\!\! \left\| L_{rw} \right\|_2 \left\|  c_1 D - I \right\|_2 \nonumber \\
\!\!\!\!\!\! &\leq& \!\!\!\! 2 \max \left( \left\lvert d_{\max} c_1 - 1 \right\rvert, \left\lvert d_{\min} c_1 -1 \right\rvert \right),\label{eq:later}\\
\!\!\!\!\!\! &=& \!\!\!\! 2 ~ \frac{d_{\max} - d_{\min}}{d_{\max} + d_{\min}},\nonumber
\end{eqnarray}
where $c_1 {\displaystyle{\EqualDef}} {2}/{(d_{\max} + d_{\min})}=1/d_1 .$
The choice for $c_1$ minimises the upper bound on the error.
So in this case the affine transformation and parameter choice have the effect
$$
L \rightarrow {\cal F}_2(L)\EqualDef\hat{L}_{rw} \approx \lrw,
$$
where ${\cal F}_2(L)$ is the affine transform specified in (\ref{eqn_defn_hat_lrw}).
The error bound $e\left( L, \lrw \right)$  satisfies 
\begin{equation}
\left\lvert \eta_i(\hat{L}_{rw}) - \eta_i \right\rvert \leq 2 ~ \frac{d_{\max} - d_{\min}}{d_{\max} + d_{\min}} 
\EqualDef
e\left( L, \lrw \right) \label{eqn_Lrw_bound}.
\end{equation}

Asymptotically, as $d_{\max} \rightarrow \infty$ with fixed $d_{\min}>0,$ the bound tends to 2. The restricted range of $\dmin$ is due to $D^{-1}$, the normalised Laplacian and therefore the bound $e(L,\lrw)$ not being defined for $\dmin =0$. As in Section \ref{sec_bd_A_L}, for $d$-regular graphs, where $d_{\max} = d_{\min} = d$, the bound equals zero. Hence, the spectra of $\hat{L}_{rw}$ and $\lrw$ are equal and from (\ref{eqn_secondmap}),
$
\eta_i=c_1\lambda_i=\lambda_i/d,
$
i.e., the eigenvalues are related by the required exact relation, as claimed in Remark~\ref{remark:dreg}.

Overall, the behaviour of this bound is similar to $e(A,L)$  obtained in Section \ref{sec_bd_A_L}. The smaller the difference of the degree sequence extremes $d_{\max}$ and $d_{\min},$ the closer the spectra of $L$ and $\lrw$ are related, as signified by a smaller bound on the eigenvalue differences.

\section{Bounding  the individual eigenvalue difference for $A$ and $L_{rw}$}\label{sec_bd_A_Lrw}

Now we move on to the final of our three possible relations: the spectra of $A$ and $L_{rw}.$ Let $\lrw$ have eigenvalues $\eta_1\leq \cdots \leq \eta_n.$ Now, with $c_2, d_2 \in \mathbb{R},$
\begin{eqnarray}
A w &=& \mu w\nonumber\\
\iff\left( d_2 I-D +L \right) w &=& \left(d_2-\mu \right)w\nonumber\\
\iff c_2 \left( d_2 I-D + D L_{rw} \right) w &=& c_2 \left(d_2-\mu \right) w \nonumber\\
\tilde{L}_{rw} w &=& \eta( \tilde{L}_{rw}) w, \label{eqn_Lrwinverted}
\end{eqnarray}
where
\begin{eqnarray}
\tilde{L}_{rw} &{\displaystyle{\EqualDef}}& c_2 \left( d_2 I -D+ D L_{rw} \right)\nonumber\\
&=&c_2(d_2I-A)\label{roleofc2} \\
\eta( \tilde{L}_{rw}) &{\displaystyle{\EqualDef}}& c_2\left(d_2-\mu\right). \label{eqn_thirdcase}
\end{eqnarray}
$\tilde{L}_{rw}$ is an affine transform of $A$ with transform parameters $c_2$ and $d_2.$ It can also be viewed as an additive and multiplicative perturbation of $\lrw.$ 
The affine transformation, followed by the parameter choices, will therefore correspond to
$$
A \rightarrow {\cal F}_3(A) \EqualDef \tilde{L}_{rw} \approx \lrw,
$$
where ${\cal F}_3(A)$ is the affine transform specified in (\ref{roleofc2}). Note again that we require $\dmin >0$ due to $\lrw$ not being well-defined for $\dmin =0$.

We now have two free parameters, $c_2$ and $d_2,$ which will be chosen  to minimise the upper bound on the eigenvalue differences.
We start with
\begin{equation}
\left\| \tilde{L}_{rw} - L_{rw}\right\|_2  \geq \left\lvert \eta_i(\tilde{L}_{rw}) - \eta_i\right\rvert=|c_2(d_2-\mu_i)-\eta_i|, \label{eqn_eval_difference_3}
\end{equation}
for $i \in \{1,2,\ldots,n\}.$
Then,
\begin{align}
\left\| \tilde{L}_{rw} - L_{rw}\right\|_2
 &= \left\| c_2 d_2 I -c_2D +  \left( c_2 D -I \right)L_{rw} \right\|_2 \nonumber\\
&\leq \left\| c_2 d_2 I-c_2D   \right\|_2 + 2 \left\| c_2 D - I \right\|_2  \label{eqn_Lrw_choice_d}\\
&\leq |c_2| \frac{d_{\max} - d_{\min}}{2}\nonumber\\
& + 2 \max\left( \left\lvert d_{\max} c_2 - 1 \right\rvert, \left\lvert d_{\min} c_2 -1 \right\rvert \right).\label{eqn_Lrw_choice_c2}
\end{align}
We choose $d_2 = (d_{\max} + d_{\min})/{2}$ exactly as in (\ref{eqn_defn_d}), since the term being minimized is identical in both cases.
The choice of $c_2$ is involved, since it appears in both terms; details are given in Appendix~\ref{app:choosec2}, where we see that 
the choice $c_2={2}/(d_{\max} + d_{\min})$
means that
\begin{equation}\label{eq:eord}
e\left( A, \lrw \right) \EqualDef  3 ~\frac{d_{\max} - d_{\min}}{d_{\max} + d_{\min}}. 
\end{equation}

For $d$-regular graphs,  the bound equals zero
and  hence the spectra of $\tilde{L}_{rw}$ and $\lrw$ are equal and from (\ref{eqn_thirdcase}),
$\eta_i=c_2(d_2-\mu_i)=1-(\mu_i/d),$
i.e.,
the spectra of $\lrw$ and $A$ are related by the required exact relation. 
\begin{remark}\label{remark:edash}
In Appendix~\ref{app:choosec2} an alternative bound, $e'\left( A, \lrw \right)$ given in (\ref{eq:edash}), is found. This is tighter, but the corresponding transformation is degenerate. We hence prefer (\ref{eq:eord}), but will also show $e'\left( A, \lrw \right)$ in visualisations in Section~\ref{sec_visualising_data}.\hfill$\square$
\end{remark}

\section{Nature of transformations}\label{sec_tranf_properties}
In order to motivate the transformations of the spectra further we will visualise them on a real data set and also analyse their properties. 
We will refer to the transformations in (\ref{eq1_Linverted}), (\ref{eqn_secondmap}) and  
(\ref{eqn_thirdcase}) as
$f_1, f_2$ and $f_3$, respectively. Here $d_1 = d_2 = {(\dmax + \dmin)}/{2}$ and $c_1 = c_2 = {2}/{(\dmax + \dmin)}$.
\begin{eqnarray}
f_1(\mu) &=& d_1-\mu; \label{eq:deffone} \\
 f_2(\lambda)&=& c_1 \lambda; \label{eq:defftwo}\\
 f_3(\mu) &=& c_2 (d_2- \mu)   = 1-c_2 \mu.\label{eq:deffthree}
\end{eqnarray}

\subsection{Karate data example}\label{subsec:firstKarate}
We first apply the transforms to the spectra of Zachary's karate dataset, which will be further analysed in Section \ref{sec_visualising_data}.
 
\begin{figure}[t!]
\begin{center}
\includegraphics[scale=0.58,clip]{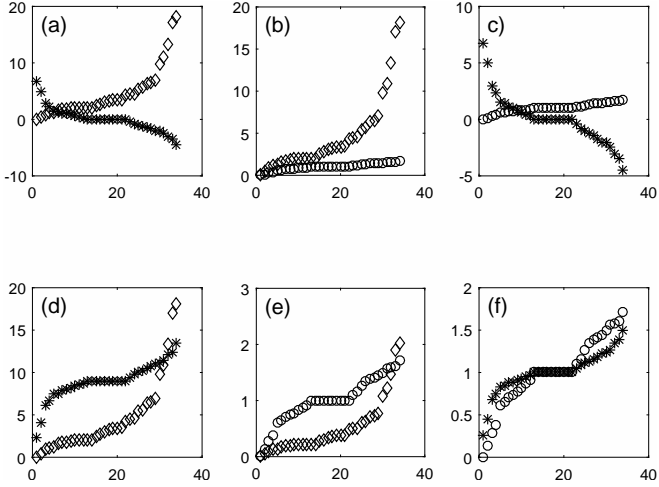}
\caption{Results for Zachary's karate dataset.
First row: (a) eigenvalues $\mu$ of $A$ (stars) and $\lambda$ of $L$  (diamonds); (b) eigenvalues $\lambda$ of $L$ and $\eta$ of $\lrw$ (circles); (c) eigenvalues $\mu$ of $A$ and $\eta$ of $\lrw$. Second row:  the first of the two eigenvalue spectra are transformed by their respective transformations $f_1, f_2$ and $f_3$.}\label{fig_transformation_visualisation_on_karate}
\end{center}
\end{figure}
The karate dataset was obtained from \cite{BatageljMrvar06} and dates back to \cite{Zachary77}.  We work with a square matrix with 34 entries, `\texttt{ZACHE},' which is the adjacency matrix of a social network describing presence of interaction in a university karate club. 
This data set is a common benchmark in the literature, as it lends itself to clustering; see, e.g., \cite{BickelPurnmarita16,ChenHero15,Fortunato10,KarrerNewman11}. 

The untransformed and transformed eigenvalues of the representation matrices $A$, $L$ and $\lrw$ of Zachary's karate dataset are shown in Fig.~\ref{fig_transformation_visualisation_on_karate}.
The eigenvalue spectra become more comparable via the proposed transformations. From Figs.~\ref{fig_transformation_visualisation_on_karate}(d), (e) and (f) it can be observed that each pair of spectra of the karate dataset cover a similar range after transformation. However,  they don't coincide and it can be argued that they carry different information about the underlying network. 

\subsection{Transformation properties}

All three transformations $f_1, f_2$ and $f_3$  are affine. The following simple observations are made for general affine transformations $g(x) = a x +b$, where $a,b \in \mathbb{R}, a\ne 0$.
\subsubsection{Ordering}
Ordering is preserved:
\begin{align*}
\mu_1 &\geq \mu_2 \\
\iff a \mu_1 +b &\geq a \mu_2 +b\\
\iff g(\mu_1) &\geq g(\mu_2).
\end{align*}
\subsubsection{Eigengaps}
Next we show that eigengaps, relative to the spectral support, are preserved. Assume that the domain of $g$ is equal to the interval $[x_1, x_2]$. Hence $g$'s image is equal to $[g(x_1),g(x_2)]$. Then normalised eigengaps are preserved:
\begin{equation}
\frac{\mu_1 -\mu_2}{x_2 -x_1} = \frac{a \left(\mu_1-\mu_2\right) +b -b}{ a \left(x_2 -x_1\right) +b -b}  = \frac{g(\mu_1) - g(\mu_2)}{g(x_2) -g(x_1)}.
\label{eq:gapspreserved}
\end{equation}

\subsubsection{Mapping of spectral supports}
As a final property we show the mappings for  the spectral supports of the  different representation matrices. The spectral supports of the three representation matrices can be derived via Gershgorin's theorem and are given in \cite[p.~29, p.~64, p.~68]{vanMieghem2011}.

\begin{eqnarray}
&& \!\!\!\!\!\! \!\!\!\!\!\!\!\!\!\!\!\! \!\!\!\!\!\!f_1: [-\dmax, \dmax] \rightarrow \left[-\frac{\dmax -\dmin}{2}, \frac{3\dmax +\dmin}{2}\right] 
\label{eq:suppone}\\
&&\!\!\!\!\!\! \!\!\!\!\!\!\!\!\!\!\!\! \!\!\!\!\!\!f_2 : [0, 2\dmax] \rightarrow \left[0,  \frac{4 \dmax}{\dmax +\dmin}\right]\label{eq:supptwo}\\
&&\!\!\!\!\!\! \!\!\!\!\!\!\!\!\!\!\!\! \!\!\!\!\!\!f_3 : \left[-\dmax, \dmax\right] \rightarrow \left[-\frac{\dmax -\dmin}{\dmax + \dmin}, \frac{3\dmax +\dmin}{\dmax + \dmin}\right].\label{eq:suppthree}
\end{eqnarray}
\begin{remark}
It is possible to choose the transformation parameters so that the transformation maps exactly to the spectral support of the target matrix.
However, this results in a greater bound on the eigenvalue differences. Rather, since the mapped supports are of little consequence, we choose our transformation parameters according to the bound value, their treatment of eigenvalue ordering, and relative eigengap preservation.\hfill$\square$
\end{remark}

\section{Relating the spectral bounds} \label{sec_relating_bounds}
We now compare the spectral bounds  with each other and illustrate their relationship. 
Firstly,
\begin{align*}
e\left(A,L\right) &\leq e\left(L,\lrw \right) \\
\iff \frac{d_{\max}-d_{\min}}{2} &\leq 2 \frac{d_{\max} -d_{\min}}{d_{\max} + d_{\min}}\\
\iff d_{\max} + d_{\min} &\leq 4.  \numberthis \label{eqn_AL_to_LLrw}
\end{align*}
Next,
\begin{align*}
e\left(A,L\right) &\leq e\left(A, \lrw \right)\\
\iff \frac{\dmax -\dmin}{2} &\leq 3 \frac{\dmax-\dmin}{\dmax +\dmin}\\
\iff \dmax +\dmin &\leq 6. \numberthis \label{eqn_AL_to_ALrw}
\end{align*}
Finally,
\begin{align*}
e\left(L,\lrw\right) &\leq e\left(A,\lrw \right)\\
\iff 2 \frac{d_{\max} -d_{\min}}{d_{\max} + d_{\min}} &\leq 3 \frac{d_{\max} -d_{\min}}{d_{\max} + d_{\min}} \numberthis \label{eqn_ALrw_to_LLrw},
\end{align*} 
which, obviously, always holds.
\begin{table*}[t]
\centering
\begin{tabular}{c c| c| c| c| c| c| c  }
\hline\hline
&&\multicolumn{6}{c}{$d_{\min}$}\\
&& 0 & 1 & 2 & 3 & 4 & 5\\
\hline
&&&&&&&\\
& 1 & ({\bf 0.5, $\boldsymbol{\cdot}$, $\boldsymbol{\cdot}$}) & ({\underline{\bf 0, 0, 0}}) & * & * & * & *\\
& 2 & ({\bf 1, $\boldsymbol{\cdot}$, $\boldsymbol{\cdot}$}) & ({\bf 0.5, 0.67, 1}) & ({\underline{\bf 0, 0, 0}}) & * & * & * \\
& 3 & ({\bf 1.5, $\boldsymbol{\cdot}$, $\boldsymbol{\cdot}$}) & (\underline{1, 1, 1.5}) & ({\tt 0.5, 0.4, 0.6}) & ({\underline{\bf 0, 0, 0}}) & * & * \\
$d_{\max}$ & 4 & (\underline{2, $\boldsymbol{\cdot}$, $\boldsymbol{\cdot}$}) & ({\tt 1.5, 1.2, 1.8}) & ({\it 1, 0.67, 1}) & (0.5, 0.29, 0.43) & ({\underline{\bf 0, 0, 0}}) & *\\
& 5 & ({\tt 2.5, $\boldsymbol{\cdot}$, $\boldsymbol{\cdot}$}) & ({\it 2, 1.33, 2}) & (1.5, 0.86, 1.29) & (1, 0.5, 0.75) & (0.5, 0.22,0.33) & ({\underline{\bf 0, 0, 0}})\\
& 6 & ({\it 3, $\boldsymbol{\cdot}$, $\boldsymbol{\cdot}$}) & (2.5, 1.43, 2.14) & (2, 1, 1.5) & (1.5, 0.67, 1) & (1, 0.4, 0.6) & (0.5, 0.18, 0.27)\\
& 7 & (3.5, $\boldsymbol{\cdot}$, $\boldsymbol{\cdot}$) & (3, 1.5, 2.25) & (2.5, 1.11, 1.67) & (2, 0.8, 1.2) & (1.5, 0.55, 0.82) & (1, 0.33, 0.5)\\
&&&&&&&\\
\hline\hline \rule{0cm}{.2cm} 
\end{tabular}
\caption{Comparing bounds on eigenvalue differences of $A, L$ and $\lrw$. The bound values are displayed as $\left(e\left(A,L\right),e\left(L,\lrw\right), e\left(A,\lrw\right)\right)$.   The 6 different regions are labelled  as follows: $e\left(A,L\right)=e\left(L,\lrw\right)= e\left(A,\lrw\right)=0$ as \underline{\textbf{bold and underlined}}, $e\left(A,L\right)<e\left(L,\lrw\right)< e\left(A,\lrw\right)$ as \textbf{bold}, $e\left(A,L\right)=e\left(L,\lrw\right)< e\left(A,\lrw\right)$ as \underline{underlined}, $e\left(L,\lrw\right) < e\left(A,L\right) < e\left(A,\lrw\right)$ as \texttt{teletyped}, $e\left(L,\lrw\right) < e\left(A,L\right)= e\left(A,\lrw\right)$ as \textit{italic} and finally  $e\left(L,\lrw\right)< e\left(A,\lrw\right)<e\left(A,L\right)$ in normal font.}
\label{tab_comparison_of_bound_values}
\end{table*}

For easier analysis of these inequalities, we display some sample values in Table~\ref{tab_comparison_of_bound_values}. The first column of Table~\ref{tab_comparison_of_bound_values} only contains values of $e(A,L)$ as for $\dmin=0$ the normalised Laplacian and hence $e(L, \lrw)$ and $e(A, \lrw)$ are not well-defined. In practice this is of little consequence since disconnected nodes are commonly removed from the dataset as a preprocessing step. 

In Table~\ref{tab_comparison_of_bound_values} the bound values are separated into 6 different regions. We start off with the  diagonal from which it is clear that, when $\dmax = \dmin,$ i.e., for $d$-regular graphs, we have a direct spectral relation.

For the $\dmax + \dmin <4$ region (\textbf{bold}),  we find that $e\left(A,L\right)$ is the smallest of the three bounds, and the spectra of $A$ and $L$ are the most closely related. We were able to anticipate this result via \eqnref{eqn_AL_to_LLrw} and \eqnref{eqn_AL_to_ALrw}.

Now for   $\dmax + \dmin = 4$ (\underline{underlined}), we find that $e\left(A,L\right) = e\left(L,\lrw\right),$ while $e\left(A,\lrw\right)$ is still larger than the other two.

From inequalities \eqnref{eqn_AL_to_LLrw} and \eqnref{eqn_AL_to_ALrw}, we expect that the ordering of the bound values changes for $\dmax + \dmin =5$ (\texttt{teletype}). The  values show us that the spectra of $A$ and $\lrw$ are now furthest apart. As expected, the spectra of $L$ and $\lrw$ have the closest relation, i.e., the smallest bound value on their eigenvalue differences. $e\left(L,\lrw\right)$ will remain the smallest bound of the three for all $\dmax + \dmin \geq 5$. 

When $\dmax + \dmin = 6$ (\textit{italic}), we have another transition point, where $e\left(A,L\right) = e\left(A,\lrw\right)$.

Finally, with $\dmax + \dmin > 6$ (normal font), we enter the last region, which will apply to the majority of graphs. Here the spectra of $L$ and $\lrw$ have the closest relation, while the bound on the eigenvalue difference of $A$ and $L$ is the largest. It is interesting to note, that since the spectral support  of neither $A,$ nor $L,$ is bounded, the bound on their eigenvalue difference is also not bounded above. This does not apply to the other two bounds as we have $e\left(L,\lrw\right)\leq 2$ and $e\left(A,\lrw\right)\leq 3$.

\section{Explaining the Structure of Table \ref{tab_comparison_of_bound_values}} \label{sec_explaining_structure_of_tab}

We  begin the forthcoming analysis with the definition of connected components in a graph.

\begin{definition} \label{def_connected_components}
A \textit{path} on a graph $G$ is an ordered list of unique vertices such that consecutive vertices are connected by an edge. A vertex set $S_k$ is called a \textit{connected component} if there exists a path between any two vertices $v_i, v_j \in S_k$ and there exists no path from any $v_i \in S_k$ to any $v_j \notin S_k$.\hfill$\square$
\end{definition}

\subsection{Partitioning}

We will illustrate the class of unweighted graphs characterised by certain degree extremes $ \dmin$ and $\dmax$ in order to visualise the class of graphs to which the bounds apply. This will allow us to explain the monotonicity of the three bounds in the sense of Table \ref{tab_comparison_of_bound_values}. We will denote the class of graphs with $\dmin = j$ and $\dmax = k$ by $\mathcal{C}_{j,k}$. For any graph, $\dmin$ and $\dmax$ are unique, and hence no graph is in two or more classes of $\{\mathcal{C}_{j,k}\}_{j=0}^{k}$. Therefore, the classes $\{\mathcal{C}_{j,k}\}_{j=0}^{k}$ are disjoint. Furthermore, since the degree extremes of a graph always exist, we have that $\{\mathcal{C}_{j,k}\}_{j=0}^{k}$ is a partition of the class of graphs with $\dmax =k,$ (denoted by $\mathcal{C}_{\cdot,k}$ in the following).

\begin{figure}[t]
\begin{center}
\includegraphics[scale=0.8]{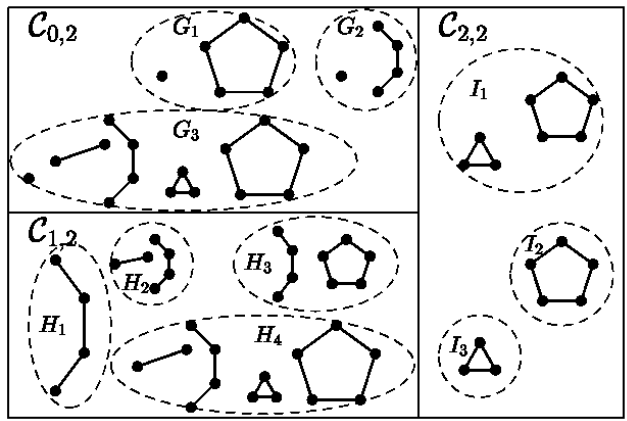}
\caption{This figure illustrates the partition $\C{\cdot}{2} = \C{0}{2} \cup \C{1}{2} \cup \C{2}{2}$. Examples of elements of the classes $\mathcal{C}_{0,2}$ containing all graphs with $\dmin =0$ and $\dmax=2$, $\mathcal{C}_{1,2}$, with element for which $\dmin =1$ and $\dmax=2$ and $\mathcal{C}_{2,2}$, the class of 2-regular graphs are shown.}
\label{fig_class_of_dmax2_graphs}
\end{center}
\end{figure}

For illustrative purposes, we consider the class of graphs $\mathcal{C}_{\cdot,2}$, i.e., the class of graphs to which the bounds in the second \textit{row} of Table \ref{tab_comparison_of_bound_values} apply. In Fig.~\ref{fig_class_of_dmax2_graphs} we display the partition of $\C{\cdot}{2},$ with examples of elements in each class. All classes $\C{j}{k}$ are infinite in size, so only a few arbitrary sample elements are displayed here. The elements of each class are marked by dashed ellipses. Graph $G_3$ is an element of $\C{0}{2}$ and consists of five connected components. Graphs in $\C{0}{2}$ are denoted by $G_a$, graphs in $\C{1}{2}$ by $H_a$ and graphs in $\C{2}{2}$ by $I_a$ for $a \in \mathbb{Z}$. 

\subsection{Adding and deleting connected components} \label{sec_relating_bounds_sub_monotonicity_rows_comp}

We now turn to a specific manipulation of graphs --- addition and removal of connected components --- which allows us to order the spectra of the graphs and hence understand the monotonicity observed in the bounds in rows of Table \ref{tab_comparison_of_bound_values}.

In Fig.~\ref{fig_class_of_dmax2_graphs}, we see that we can obtain $G_3$ from $H_4$ by adding a single disconnected vertex and that we can obtain $H_4$ from $I_1$ by adding the connected components, the line and the 2-complete component. In general, we can obtain a graph in $\mathcal{C}_{j,k}$ from a graph in $\mathcal{C}_{j+1,k},$ for all $j\leq k-1, k \in \mathbb{N},$ by adding  to the graph one or more connected components
 in which all vertices are of degree greater or equal to $j$ and smaller or equal to $k,$ with at least one vertex attaining degree $j$. 

\subsection{Analysis of bound monotonicity in  rows of Table~\ref{tab_comparison_of_bound_values}}\label{sec_relating_bounds_sub_monotonicity_rows}

As stated in \cite[p.~7]{Chung97}, the spectrum of a graph is the union of the spectra of its connected components, i.e., the union of the sets of eigenvalues of the representation matrices corresponding to the connected components. 
Since all graphs in $\mathcal{C}_{j,k}$ can be obtained from graphs in $\mathcal{C}_{j+1,k}$ by adding one or more connected components,
 it can be argued that all spectra of graphs in $\mathcal{C}_{j+1,k}$ are subsets of spectra of graphs in $\mathcal{C}_{j,k}$. Therefore, the support of the spectra of graphs in $\mathcal{C}_{j,k}$ must be larger or equal to the support of spectra of graphs in $\mathcal{C}_{j+1,k}$. Hence, we expect the spectral bounds we derived to be decreasing or constant with increasing $\dmin$ and constant $\dmax$. This phenomenon can be observed in Table~\ref{tab_comparison_of_bound_values} when traversing each {\it row}. 
 
\subsection{Analysis of bound monotonicity in columns of Table~\ref{tab_comparison_of_bound_values}} \label{sec_relating_bounds_sub_monotonicity_cols}
In  Section \ref{sec_relating_bounds_sub_monotonicity_rows_comp}, when 
traversing the rows of Table \ref{tab_comparison_of_bound_values}, we were adding connected components to graphs in  $\mathcal{C}_{j+1,k}$ to obtain graphs in  $\mathcal{C}_{j,k}$. We observed decreasing bound size with increasing indices corresponding to $\dmin$. Here we  find increasing bound size with increasing indices corresponding to $\dmax:$  it seems sensible that increasing the support of the degree distribution should also increase the spectral support and hence the bounds. (The \textit{degree distribution} is a probability distribution from which vertex degrees of a graph are sampled.)

We argue that any graph in  $\mathcal{C}_{j,k+1},$ can be obtained from a graph in  $\mathcal{C}_{j,k},$ by adding a connected component with all vertex degrees greater or equal to $j$ and smaller or equal than $k+1$ with at least one node attaining degree $k+1$. Then the argument goes exactly as in Section \ref{sec_relating_bounds_sub_monotonicity_rows}, that spectra of graphs in $\mathcal{C}_{j,k}$ are subsets of spectra of graphs $\mathcal{C}_{j,k+1}$ and therefore the spectral support and hence the spectral bounds on $\mathcal{C}_{j,k+1}$ have to be greater or equal than the respective quantities for $\mathcal{C}_{j,k}$. Again, this kind of monotonicity is observed when comparing bound values in the {\it columns}\/ of Table \ref{tab_comparison_of_bound_values}.

\section{Visualising the eigenvalue bounds on data} \label{sec_visualising_data}
In this section we illustrate the bounds on the spectra of Zachary's karate dataset, which we already met in Section~\ref{subsec:firstKarate}. We then proceed to explore the bounds on three different graphs with $\dmin = 1$ and $\dmax = 17,$  choices appropriate for the karate dataset. 

\begin{figure}[t]
\begin{center}
\includegraphics[scale=0.59,clip]{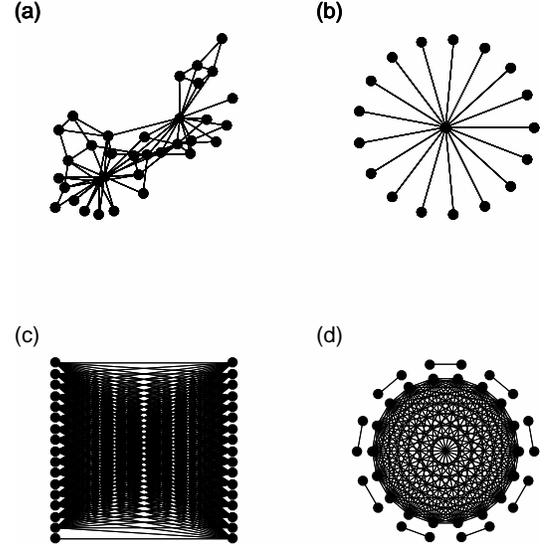}
\caption{(a) graph given by the karate data set. (b) graph $A$, the star graph on 18 nodes. (c) graph $B$, a bipartite graph with degree distribution $\{1, \{16\}^{16}, \{17\}^{17}\}$ (see Section \ref{sec_visualising_data_3examples}). (d) graph $C$, a graph consisting of 9 2-complete components and one 18-complete component.}
\label{fig_graph_plots}
\end{center}
\end{figure}

The graph defined in the karate data set is plotted in Fig.~\ref{fig_graph_plots}(a) together with the three graph examples in Figs.~\ref{fig_graph_plots}(b)-(d) analysed in Section \ref{sec_visualising_data_3examples}.

\subsection{Zachary's karate dataset} \label{sec_visualising_data_karate}

\begin{figure}[t]
\begin{center}
\includegraphics[scale=0.59,clip]{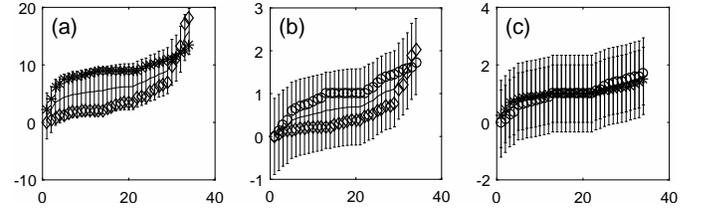}
\caption{Eigenvalue bounds on the karate eigenvalues. In plot (a) we display the bound $e(A,L)$ via the intervals together with the transformed eigenvalues of the adjacency matrix $f_1(\mu)$ (stars) and the eigenvalues of the Laplacian $\lambda$ (diamonds).  (b) the bound $e(L,\lrw)$ is displayed via the intervals, the diamonds correspond to the transformed Laplacian eigenvalues $f_2(\lambda)$ and the circles are the eigenvalues of the  normalised graph Laplacian $\eta$. In plot (c) the inner bounds interval corresponds to $e'(A,\lrw)$, (see Remark~\ref{remark:edash}), while the outer interval corresponds to the bound $e(A,\lrw)$, the stars correspond to the transformed adjacency eigenvalues $f_3(\mu)$ and the circles are the  normalised Laplacian eigenvalues $\eta$.}
\label{fig_karate_bounds_1}
\end{center}
\end{figure}

In Fig.~\ref{fig_karate_bounds_1}, we display a proof of concept of the bounds derived in this document. We display the transformed eigenvalues together with the eigenvalues we compare to and the derived bounds. The eigenvalue bounds are centred around the average value of each eigenvalue pair in order to highlight the maximal difference achievable by each individual eigenvalue pair under comparison. For the karate dataset the bound values $(e(A,L), e(L,\lrw), e(A,\lrw))$ are equal to $(8.00,1.78,2.67)$. With $\dmin = 1$ and $\dmax = 17$ these bounds clearly fall into the region characterised last in Section \ref{sec_relating_bounds}, for which $\dmax + \dmin >6$. 
The values in plot (a) are much larger than the ones in plots (b) and (c), as was to be expected from the analysis in Section \ref{sec_relating_bounds}. The particular bounds displayed here are valid for all graphs with $\dmin = 1$ and $\dmax = 17,$ i.e., all graphs in ${\cal C}_{1,17}.$ The fact of the bounds  being almost attained in plot (a), and not attained in plot (c), is more  a consequence of  the structure of the graph given by the karate data set than tightness and quality of the bounds. Since the three bounds $e(A,L)$, $e(L, \lrw)$ and $e(A, \lrw)$ apply to entire classes $\C{j}{k}$ at a time, we can only hope to achieve tightness on these classes (hence the bounds being attained for some elements in $\C{j}{k}$), and not on each individual element of them.

\subsection{Three examples exploring the bounds for graphs in $\C{1}{17}$} \label{sec_visualising_data_3examples}

We now proceed to explore the class of graphs with $\dmin=1$ and $\dmax=17$, $\mathcal{C}_{1,17}$, to which the bounds for the karate dataset displayed in Fig.~\ref{fig_karate_bounds_1} apply. We will consider the star on 18 vertices, ``graph A,'' a bipartite graph, ``graph B,'' and a graph containing several 1- and 17-regular connected components, ``graph C.'' 

\subsubsection{Graph A (Star graph)}
\begin{figure}[t!]
\begin{center}
\includegraphics[scale=0.59,clip]{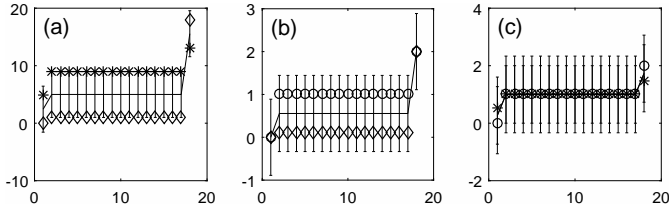}
\caption{Spectra and transformed spectra with their bound for graph A, a star on 18 nodes, with degree sequence $\{\{1\}^{17},17\}$. The format follows Fig.~\ref{fig_karate_bounds_1}.}
\label{fig_karate_bounds_star}
\end{center}
\end{figure}

Fig.~\ref{fig_karate_bounds_star} displays the three bounds on the spectra of the star graph with degree sequence $\{\{1\}^{17},17\}$, where the notation $\{x\}^{y}$ is shorthand for the multiset consisting of $y$ elements equal to $x$.

Especially notable, is that from Fig.~\ref{fig_karate_bounds_star}(a) the bound $e(A,L)$ can be seen to be tight: the spectra of $A$ and $L$ have maximal distance on individual eigenvalue level for 16 of the 18 eigenvalues. Further,
for the 16 eigenvalues of maximal distance between spectra of $A$ and $L,$ the spectra of $A$ and $\lrw$ coincide, up to transformation.

\subsubsection{Graph B (Bipartite graph)}
We will now work with a bipartite graph on 34 nodes. The degree sequence of the bipartite graph is $\{1, \{16\}^{16}, \{17\}^{17}\}$.  The graphs falling under the same spectral bound are only restricted in their degree sequence extremes, not by the number of nodes in the graphs.

\begin{figure}[t]
\begin{center}
\includegraphics[scale=0.59,clip]{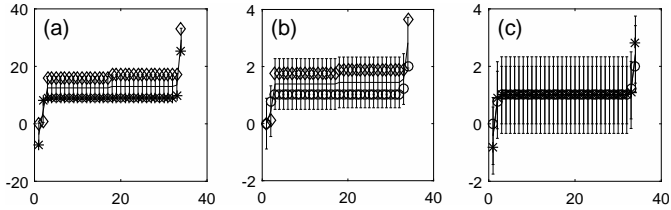}
\caption{Spectra and transformed spectra with their bound for graph B, the bipartite graph on 34 nodes, with degree sequence $\{1, \{16\}^{16}, \{17\}^{17}\}$. The format follows Fig.~\ref{fig_karate_bounds_1}.}
\label{fig_karate_bounds_bipartite_1}
\end{center}
\end{figure}

In Fig.~\ref{fig_karate_bounds_bipartite_1}(a), 15 of the eigenvalue pairs attain the maximal distance given by the bound and the  remaining 19 eigenvalue pairs are  close to attaining the maximal distance. Also, all but one of the eigenvalues of the graph Laplacian $L$ are  larger than the ones of the adjacency matrix $A$. Fig.~\ref{fig_karate_bounds_star} displayed the opposite phenomenon, the eigenvalues of $A$ being mostly larger than the eigenvalues of $L$. Another, very interesting, phenomenon can be spotted in the behaviour of the first two eigenvalues of $A$ and $L$ in Fig.~\ref{fig_karate_bounds_bipartite_1}(a). To discuss this we need the following:

\begin{definition} \label{def_max_crossover}
Let $X$ and $Y$ be real symmetric matrices with eigenvalues $\lambda_1(X), \ldots, \lambda_n(X)$ and $\lambda_1(Y), \ldots, \lambda_n(Y)$, respectively and let $\phi$ be an affine transformation applied to a matrix spectrum. Then, in the context of a bound of the form $h(X, Y) \geq \left\lvert \phi(\lambda_i(X)) - \lambda_i(Y) \right\rvert,$ we will use the term \textit{maximal crossover} of eigenvalues to express that $\phi(\lambda_i(X))$ attains one of the ends of the bound $h(X, Y)$, $\phi(\lambda_{i+1}(X))$ attains the opposite end of the bound $h(X, Y),$ and  vice versa for $\lambda_i(Y)$ and $\lambda_{i+1}(Y)$. \hfill$\square$
\end{definition}

We see that the first two eigenvalues of $A$ and $L$ are very close to a maximal crossover, where $\phi=f_1,$ with rounded distances $\lambda_1 - f_1(\mu_1) = 7.49$ and $\lambda_2 - f_1(\mu_2) = -7.06$ for a bound value $e(A,L) = 8$. The concept of a maximal crossover will be significant in Section \ref{sec_eigengap_bound} when bounding eigengaps.

Fig.~\ref{fig_karate_bounds_bipartite_1}(b) also provides an interesting insight:  the last two eigenvalues $f_2(\lambda_{34})$ and $\eta_{34}$ attain the opposing ends of the bound $e(L,\lrw),$ with a rounded distance between them equal to $f_2(\lambda_{34}) - \eta_{34}= 1.67$ on the bound value $e(L,\lrw) = 1.78$. Hence, $e(L,\lrw)$ is somewhat close to  tight on 
${\cal C}_{1,17}.$ 

Finally, we see in Fig.~\ref{fig_karate_bounds_bipartite_1}(c) that the eigenvalues of $A$ and $\lrw$ mostly coincide up to transformation, as  was the case for the eigenvalues in  Fig.~\ref{fig_karate_bounds_star}(c). However, as mentioned previously, this has no particular implication for  the tightness of the bound, as the bound applies to a very large class of graphs.

In both Figs.~\ref{fig_karate_bounds_star} and \ref{fig_karate_bounds_bipartite_1}, the difference between the transformed spectrum of the adjacency matrix and the spectrum of the normalised Laplacian is notably smaller than the differences in the other two spectral comparisons. A possible explanation of this phenomenon follows from noticing that the difference between the adjacency matrix $A$ and the normalised Laplacian $\lrw = I-D^{-1}A$ is the degree normalisation. It is hence conceivable, that for degree sequences with low `variance,' this normalisation is less impactful and hence we expect a small difference in the spectra. This `low variance' argument is a possible explanation for the small differences observed in Figs.~\ref{fig_karate_bounds_star}(c) and \ref{fig_karate_bounds_bipartite_1}(c). This phenomenon positively supports our transformations, specifically $f_3,$ through recovery of the small difference between the two spectra.

\subsubsection{Graph C}
We can easily construct an example for which the spectra $f_3(\mu)$ and $\eta$ show larger differences by noticing that the value $c_2$ in the transformation function $f_3$ is fixed to be equal to $1/9$ for all graphs in $\C{1}{17}$. As we saw in Section \ref{sec_relating_bounds}, $\C{1}{17}$ also contains graphs which consist of several connected components. We hence consider the graph consisting of 9 pairs of vertices connected only to one another, hence 18 nodes of degree 1, and a complete connected component on 18 nodes. 
 
\begin{figure}[t]
\begin{center}
\includegraphics[scale=0.59,clip]{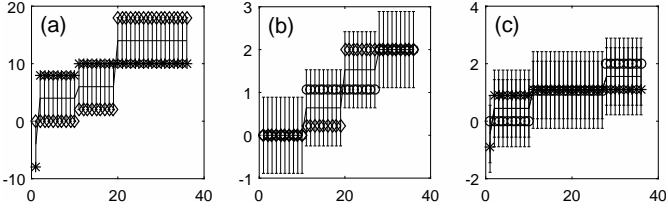}
\caption{Spectra and transformed spectra together with their bound for graph C, consisting of 10 connected components, 9 of which are the complete graph on 2 nodes and one of which is the complete graph on 18 nodes. Therefore the degree sequence is equal to $\{ \{1\}^{18},\{17\}^{18}\}.$ The format follows Fig.~\ref{fig_karate_bounds_1}.}
\label{fig_karate_bounds_disconnected}
\end{center}
\end{figure}

Consider Fig.~\ref{fig_karate_bounds_disconnected}(c): for just the 1-regular components, choosing the scaling to be equal to $c_2^{(1)} = 1$ would yield perfect correspondence of the spectrum of the transformed adjacency matrix and the spectrum of the normalised graph Laplacian ($\lrw$). For just the 17-regular component, the normalisation by $c_2^{(2)} = 1/17$ is ideal. The choice of $c_2=1/9$ yields a difference in the transformed spectrum and the untransformed spectrum, since it represents a departure from the case of exact correspondence of spectra. There is a larger difference between the eigenvalues corresponding to the complete graphs on two nodes, since the departure of the scaling factor $c_2$ from the optimal scaling factor $c_2^{(1)}$ is larger than for the eigenvalues corresponding to the 17-regular component. Correspondence of eigenvalues to different connected components of the graph could be established by observing connected components {\it separately}, in conjunction with the result in \cite{Chung97} discussed in Section \ref{sec_relating_bounds_sub_monotonicity_rows}.

Finally, in Fig.~\ref{fig_karate_bounds_disconnected}(a), we spot a maximal crossover with $\phi = f_1$ between eigenvalues $f_1(\mu_1)$, $\lambda_1$ and $f_1(\mu_2)$, $\lambda_2$ and again for the pairs with indices 19 and 20.

\begin{remark}
In the case of a graph consisting of more than one connected component, as in Fig.~\ref{fig_karate_bounds_disconnected}, it is beneficial to consider connected components separately. When this is done, the degree sequence extremes are decreased or stay equal and hence the transformation becomes more accurate or stays the same and the bounds decrease or stay equal (see Sections~\ref{sec_relating_bounds_sub_monotonicity_rows} and \ref{sec_relating_bounds_sub_monotonicity_cols}). For graph C underlying Fig.~\ref{fig_karate_bounds_disconnected}, the spectra of the connected components can individually be exactly related as the connected components are  1- and 17-regular.\hfill$\square$
\end{remark}

\section{Bounds on Normalised eigengap differences} \label{sec_eigengap_bound}

Here we examine the eigengaps corresponding to the three representation matrices. We will compare bounds on eigengaps normalised by the spectral support, to bounds on eigengaps derived from eigenvalues transformed by $f_1, f_2$ and $f_3$. Furthermore, we investigate under which conditions on the spectra the derived eigengap bounds  are tight.

\subsection{Normalised eigengap difference of $A$ and $L$} \label{sec_eigengap_bound_A_L}

Let $\mathcal{M}_i $ denote the $i^{\mathrm{th}}$ eigengap of $A$, $\mathcal{M}_i = \mu_i - \mu_{i+1}$ and  ${\mathcal L}_i$ denote the $i^{\mathrm{th}}$ eigengap of $L$, ${\mathcal L}_i = \lambda_{i+1}-\lambda_i$ for $i \in \{1,2,\ldots,n-1\}$. Also recall that the spectral support of $A$ and $L$ is equal to $[-\dmax, \dmax]$ and $[0, 2 \dmax],$ respectively, so the length of the support is  
$\lsupp(\mu)=\lsupp(\lambda)= 2 \dmax.$

From (\ref{eq:deffone}) and (\ref{eq:suppone}), 
\begin{eqnarray*}
&&\!\!\!\!\!\!\!\left\lvert \frac{f_1(\mu_{i+1}) - f_1(\mu_i)}{\lsupp(f_1(\mu))} - \frac{{\mathcal{L}}_i}{\lsupp(\lambda)} \right\rvert= \left\lvert \frac{\mathcal{M}_i}{2 \dmax} - \frac{{\mathcal L}_i}{2 \dmax} \right\rvert \\
&=& \left\lvert \frac{\mathcal{M}_i}{\lsupp(\mu)} - \frac{{\mathcal L}_i}{\lsupp(\lambda)} \right\rvert,
\end{eqnarray*}
which would be anticipated from (\ref{eq:gapspreserved}). Furthermore, $\lsupp(\mu)=\lsupp(\lambda),$ so the normalisation of the two bounds separately is equivalent to normalising the entire difference by a single value. This will not be the case in Sections~\ref{sec_eigengap_bound_L_Lrw} and \ref{sec_eigengap_bound_A_Lrw}, where the lengths of the eigenvalue supports are different.

The bound on the normalised eigengap difference, denoted $g(A,L),$ then takes the following form:
\begin{align*}
 \left\lvert \frac{\mathcal{M}_i}{2 \dmax} - \frac{{\mathcal L}_i}{2 \dmax} \right\rvert &= \frac{1}{2\dmax} \left\lvert \mu_i - \mu_{i+1} -\lambda_{i+1} +\lambda_i \right\rvert\\
& = \frac{1}{2\dmax} \left\lvert \mu_i - \frac{\dmax + \dmin}{2}\right. \\
&\quad\left.+\lambda_i - \mu _{i+1} + \frac{\dmax + \dmin}{2} -\lambda_{i+1}  \right\rvert\\
& = \frac{1}{2\dmax} \left\lvert \left[ \lambda_i - f_1(\mu_i) \right]\right.\\
& \quad\left. + \left[ f_1(\mu _{i+1}) -\lambda_{i+1}\right] \right\rvert \numberthis \label{eqn_eigengaps_AL_tight}\\
& \leq \frac{1}{2\dmax} 2 e(A,L) \numberthis\label{eq:ALtwo} \\
& = \frac{\dmax - \dmin}{2 \dmax} \EqualDef g(A,L).
\end{align*}
The triangle inequality has been used to go from (\ref{eqn_eigengaps_AL_tight}) to (\ref{eq:ALtwo}).

It follows from Equation \eqnref{eqn_eigengaps_AL_tight}, that the bound on the eigengaps is tight if and only if the individual mapped eigenvalue differences $\lambda_i - f_1(\mu_i) $ and $ \lambda_{i+1} - f_1(\mu _{i+1})$ have opposing signs and occupy the extremes of the bound $e(A,L)$. 
In Definition~\ref{def_max_crossover}, such a situation was called a maximal crossover for general transformations $\phi$. We observed in Section  \ref{sec_visualising_data}, Fig.~\ref{fig_karate_bounds_bipartite_1}(a), a situation very close to a maximal crossover with $ \phi =f_1.$ 
A maximal crossover is shown in
Fig.~\ref{fig_karate_bounds_disconnected}(a)  for $\phi=f_1;$ therefore, $g(A,L)$ is the optimal {\it constant}\/ bound since it is tight for some cases in  $\C{1}{17}.$

\subsection{Normalised eigengap difference of $L$ and $\lrw$}
\label{sec_eigengap_bound_L_Lrw}
We proceed to derive a bound on the normalised eigengap difference of $L$ and $\lrw$, denoted $g(L, \lrw)$. Let $\mathcal{N}_i = \eta_{i+1}-\eta_i$ denote the $i^{\mathrm{th}}$ eigengap of $\lrw$ for $i \in \{1, \ldots, n-1\}.$ For $\lrw$ we recall that $\lsupp(\eta_i)=2$.

Firstly, when normalising the transformed eigenvalues by their support, we end up with the comparison of normalised untransformed spectra as proven in Section~\ref{sec_tranf_properties}, i.e.,
from (\ref{eq:defftwo}) and (\ref{eq:supptwo}),
\begin{align*}
\left\lvert \frac{ f_2(\lambda_{i+1}) \!-\! f_2(\lambda_i)}{ \lsupp(f_2(\lambda))} \!- \!\frac{\mathcal{N}_i}{\lsupp(\eta)} \right\rvert \!&= \!\left\lvert \frac{c_1}{\frac{2}{\dmax +\dmin}} \frac{ [\lambda_{i+1} - \lambda_i]}{ 2 \dmax} - \frac{\mathcal{N}_i}{2} \right\rvert \\
\!&=\! \left\lvert  \frac{ {\mathcal L}_i}{\lsupp(\lambda)} - \frac{\mathcal{N}_i}{\lsupp(\eta)} \right\rvert,
\end{align*}
as expected from (\ref{eq:gapspreserved}).
Now the normalised eigengap difference follows:
\begin{eqnarray}
\left\lvert \frac{{\mathcal L}_i}{2 \dmax} - \frac{\mathcal{N}_i}{2 } \right\rvert \!\!\!&=&\!\!\!\! \frac{1}{2} \left\lvert \frac{1}{\dmax}\left(\lambda_{i+1} - \lambda_i \right)+ \eta_i - \eta_{i+1} \right\rvert\label{eq:nigone}\\
\!\!\!&\leq& \!\!\!\!\!\frac{1}{2} \left[\left\lvert \frac{1}{\dmax} \lambda_{i+1} - \eta_{i+1}\right\rvert + \left\lvert \frac{1}{\dmax} \lambda_i - \eta_i \right\rvert\right],\nonumber
\end{eqnarray}
from the triangle inequality.
A comparison with (\ref{eq:equiv}), shows that
these eigenvalue differences can be bounded by replacing $c_1$ in (\ref{eq:equiv}) by  $ {1}/{\dmax}$ and then making this substitution into (\ref{eq:later}):

\begin{align*}
\left\lvert \frac{{\mathcal L}_i}{2 \dmax} - \frac{\mathcal{N}_i}{2 } \right\rvert & \leq \frac{1}{2}\left[ 4 \max\left(0, \left\lvert \frac{\dmin}{\dmax} - 1 \right\rvert \right) \right]\\
&= 2 \left( 1 - \frac{\dmin}{\dmax}\right)\\
&= 2 \frac{\dmax -\dmin}{\dmax} \EqualDef  g(L, \lrw).
\end{align*}

We see from (\ref{eq:nigone}), that observing a maximal crossover with $\phi(\lambda) = \lambda/\dmax$ in the normalised spectra, would render $g(L,\lrw)$ a tight bound.

\begin{remark}
Suppose we ignore the fact that the spectral support of the eigenvalues is not equal and just compare transformed eigengaps $f_2(\lambda_{i+1}) - f_2(\lambda_i)$ to the eigengaps of $\lrw$. (We still normalise by $
1/\lsupp(\eta)=1/2$ to have the spectrum of the normalised graph Laplacian have the same support, as in (\ref{eq:nigone}).)
Then we have
\begin{eqnarray*}
&&\!\!\!\!\!\!\!\!\!\!\!\!\!\!\!\!\!\!\frac{1}{\lsupp(\eta)} \left\lvert f_2(\lambda_{i+1}) - f_2(\lambda_i) - \mathcal{N}_i \right\rvert \\
&&\leq \frac{1}{2} \left[\left\lvert f_2(\lambda_{i+1}) - \eta_{i+1}\right\rvert + \left\lvert f_2(\lambda_i) - \eta_i \right\rvert\right]\\
&&\leq e(L,\lrw) = 2 \frac{\dmax - \dmin}{\dmax + \dmin}  \EqualDef  g'(L, \lrw).
\end{eqnarray*}
So, by relaxing the restriction that the eigenvalues should be compared on the same scale, and by normalising by 1/2, we obtained a smaller bound on the transformed, but unnormalised, eigengaps, i.e., $g'(L, \lrw) \leq g(L,\lrw)$. However, when obtaining $g'(L,\lrw),$ we compare transformed eigenvalues $f_2(\lambda)/2 \in [0, 2 {\dmax}/{(\dmax +\dmin)}]$ to eigenvalues $\eta/2 \in [0,1]$. This can be argued to be less intuitive from a practitioners point of view and less well motivated in application. We hence recommend the comparison of eigengaps normalised by the eigenvalue's spectral support and the bound $g(L, \lrw)$ which results from this comparison. \hfill$\square$
\end{remark}

\subsection{Normalised eigengap difference of $A$ and $\lrw$} \label{sec_eigengap_bound_A_Lrw}
Finally, we will derive the bound on the normalised eigengaps of $A$ and $\lrw,$ denoted by $g(A, \lrw)$. 
From (\ref{eq:deffthree}) and (\ref{eq:suppthree}), we can see that (\ref{eq:gapspreserved}) holds:
\begin{align*}
\left\lvert \frac{ f_3(\mu_{i+1}) - f_3(\mu_i)}{ \lsupp(f_3(\mu))} - \frac{\mathcal{N}_i}{\lsupp(\eta)} \right\rvert &= \left\lvert \frac{c_2}{\frac{2}{\dmax +\dmin}} \frac{ \mu_i - \mu_{i+1}}{ 2 \dmax} - \frac{\mathcal{N}_i}{2} \right\rvert \\
&= \left\lvert  \frac{ \mathcal{M}_i}{\lsupp(\mu)} - \frac{\mathcal{N}_i}{\lsupp(\eta)} \right\rvert.
\end{align*}

We hence proceed to find the bound on the normalised eigengaps.

\begin{align*}
\left\lvert \frac{\mathcal{M}_i}{2 \dmax} \!-\! \frac{\mathcal{N}_i}{2} \right\rvert & = \frac{1}{2} \left\lvert \frac{1}{\dmax}\left(\mu_{i} - \mu_{i+1} +d_2 - d_2 \right)+ \eta_i - \eta_{i+1} \right\rvert\\
&\leq \frac{1}{2}\left[ \left\lvert \frac{1}{\dmax} (d_2-\mu_{i+1}) - \eta_{i+1} \right\rvert\right. \\
&\quad\left.+ \left\lvert \frac{1}{\dmax} (d_2-\mu_i)  - \eta_i \right\rvert\right],
\end{align*}
via the triangle inequality.
Since the optimisation of $d_2$ is independent of the scaling factor in (\ref{eqn_Lrw_choice_d}), we find again that $d_2 = {(\dmax + \dmin)}/{2}$ minimises the bound. 
Then, a comparison with (\ref{eqn_eval_difference_3}), shows that these eigenvalue differences can be bounded by replacing $c_2$ by ${1}/{\dmax}$ and then substituting into (\ref{eqn_Lrw_choice_c2}) to give
\begin{align*}
\left\lvert \frac{\mathcal{M}_i}{2 \dmax} - \frac{\mathcal{N}_i}{2 } \right\rvert & \leq \frac{\dmax -\dmin}{2 \dmax} + 2 \max\left(0, \left\lvert \frac{\dmin}{\dmax} -1 \right\rvert \right)\\
&= \frac{5}{2} \frac{(\dmax -\dmin)}{ \dmax} \EqualDef  g(A, \lrw).
\end{align*}
Again, as in Sections \ref{sec_eigengap_bound_A_L} and \ref{sec_eigengap_bound_L_Lrw}, a maximal crossover 
with $\phi(\mu)=(d_2-\mu)/\dmax$
in the normalised spectra corresponds to $g(A,\lrw)$ being a tight bound.

\begin{remark}
Suppose, as in Section~\ref{sec_eigengap_bound_L_Lrw}, we relax the normalisation restriction that both spectra have to be on the same scale after normalisation and compare the transformed eigenvalue gaps to the eigengaps of $\lrw$ both normalised by $1/2$:
\begin{align*}
{\textstyle\frac{1}{2}} \left\lvert f_3(\mu_{i+1}) - f_3{\mu_i} - \mathcal{N}_i  \right\rvert 
&\leq {\textstyle{\frac{1}{2}}} \left[\left\lvert f_3(\mu_{i+1}) - \eta_{i+1}\right\rvert\right.\\ 
&\left.\quad+ \left\lvert f_3(\mu_i) - \eta_i \right\rvert \right]\\
&\leq e'(A,\lrw) \EqualDef  g'(L, \lrw) \\
&= \begin{cases}
 3 ~\frac{d_{\max} - d_{\min}}{d_{\max} + d_{\min}}, & d_{\max} \leq 5 d_{\min};\\
2, & \text{otherwise.}
\end{cases}
\end{align*}
Although $g'(A, \lrw) \leq g(A, \lrw),$ as in Section~\ref{sec_eigengap_bound_L_Lrw}, we advise against $g'(A,\lrw)$ and in favour of $g(A, \lrw)$. From a practitioners point of view it is more sensible to compare eigengaps on equal support.\hfill$\square$
\end{remark}

\section{Eigengap bound examples for varying degree extremes} \label{sec_eigengap_bound_examples}
In this section we observe our eigengap bounds on examples with varying degree extremes. The functional form of our bounds implies that a growing degree extreme difference allows for a greater spectral difference between graph representation matrices. We visualise this effect by varying 
the degree maximum in graph C by replacing the $18$-complete component by a $k$-complete component, the resulting graphs are denoted $C(k)$. The graph C we have considered so far will hence be referred to as graph $C(18)$ in the new notation.

\begin{figure}[t]
\begin{center}
\includegraphics[scale=0.59,clip]{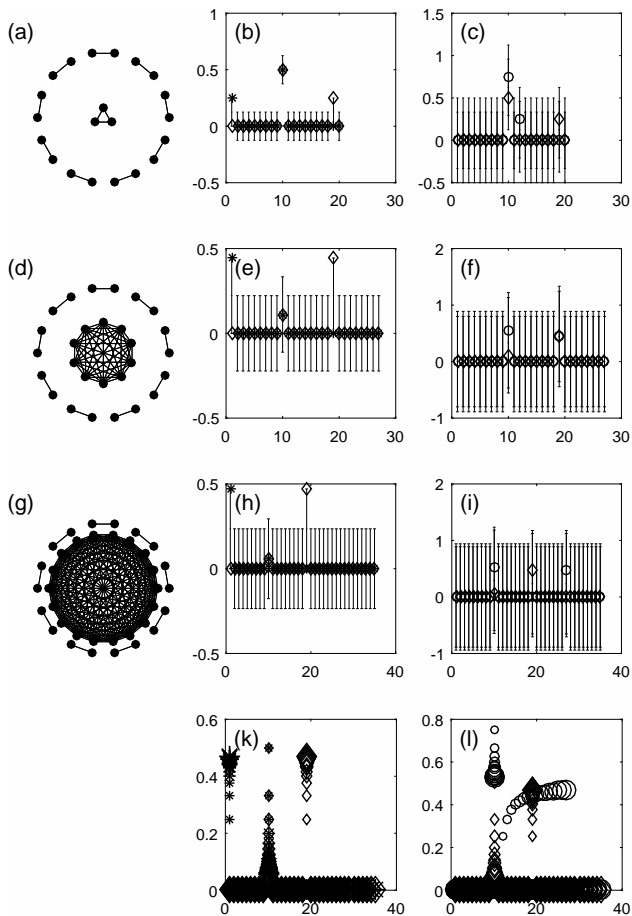}
\caption{Effects of changing $\dmax$ by varying the size of the $k$-complete component in graphs $C(k)$ with $k\in \{3, \ldots, 18\}$. (a), (d) and (g) contain plots of graphs $C(3)$, $C(10)$ and $C(18)$, respectively. (b), (e) and (h) bound $g(A,L)$ (shown via the intervals), together with the normalised eigengaps of the adjacency matrix $\mathcal{M}_i/\lsupp(\mu)$  (stars) and the normalised eigengaps of the Laplacian $\mathcal{L}_i/\lsupp(\lambda)$ (diamonds) corresponding to the graphs displayed at the start of their respective row of plots. (c), (f) and (i) bound $g'(L,\lrw)$ (shown via inner intervals), bound $g(L,\lrw)$ (via outer intervals), normalised eigengaps of the Laplacian  $\mathcal{L}_i/\lsupp(\lambda)$ (diamonds) and normalised eigengaps of the  normalised  Laplacian $\mathcal{N}_i/\lsupp(\eta)$ (circles). (k) normalised eigengaps of the adjacency matrices (stars) and the normalised eigengaps of the Laplacian (diamonds) corresponding to graphs $C(k)$ with $k\in \{3, \ldots, 18\}$. (l) normalised eigengaps of the Laplacian (diamonds) and the normalised eigengaps of the normalised Laplacian (circles) corresponding to graphs $C(k)$ with $k\in \{3, \ldots, 18\}$. In plots (k) and (l) the marker size of the normalised eigengaps increases with growing $k$.}
\label{fig_eigengap_bounds}
\end{center}
\end{figure}

The effects of changing $\dmax$ by varying the size of the $k$-complete component in graphs $C(k)$ with $k\in \{3, \ldots, 18\}$ are shown in Fig.~\ref{fig_eigengap_bounds}. Figs.~\ref{fig_eigengap_bounds}(a), (d) and (g) display plots of graphs $C(3)$, $C(10)$ and $C(18)$, respectively. Figs.~\ref{fig_eigengap_bounds}(b), (e) and (h) show the eigengap bounds on the adjacency and unnormalised Laplacian normalised eigengaps corresponding to graphs $C(3)$, $C(10)$ and $C(18)$, respectively. Figs.~\ref{fig_eigengap_bounds}(c), (f) and (i) show the two derived bounds (see Section \ref{sec_eigengap_bound_L_Lrw}) on the unnormalised and normalised Laplacian normalised eigengaps. Figs.~\ref{fig_eigengap_bounds}(k) and (l) highlight the evolution of the eigengaps of the representation matrices as we vary $k$ from 3 to 18; the eigengap markersize was chosen to increase with $k$ to clearly show the evolution of the normalised eigengaps. 

Note that the comparison of magnitudes of eigengaps corresponding to different representation matrices in Fig.~\ref{fig_eigengap_bounds} is sensible since we normalised all eigengaps by their spectral support and hence have all eigengaps on the same scale.

While significant differences in the spectra of $A$ and $L$ were observed in Fig.~\ref{fig_karate_bounds_disconnected}(a) of Section~\ref{sec_visualising_data_3examples}, the eigengaps  mostly agree (see Fig.~\ref{fig_eigengap_bounds}(h)). Noticeably larger than other eigengaps are the first eigengap of the adjacency matrix, the $10^{\mathrm{th}}$ eigengap in both spectra and $19^{\mathrm{th}}$ eigengap in the unnormalised Laplacian spectrum. The spectra of $A$ and $L$ agree on the $10^{\mathrm{th}}$ eigengap, which is found to decrease with increasing $\dmax$ in Fig.~\ref{fig_eigengap_bounds}(k).  The first and $19^{\mathrm{th}}$ eigengaps show clear disagreement between representation matrix spectra, and for these eigengaps the bound $g(A,L)$ can be seen to be tight on all three displayed examples, graphs $C(3)$, $C(10)$ and $C(18)$. The tightness of the bound here is due to the maximal crossover discussed in Section \ref{sec_eigengap_bound_A_L}. Fig.~\ref{fig_eigengap_bounds}(k) shows the difference in the eigengaps of the two representation matrices growing with increasing degree extreme difference.

Turning now to Figs.~\ref{fig_eigengap_bounds}(c), (f) and (g) it is seen that the difference between the two proposed bounds $g(L,\lrw)$ and $g'(L,\lrw)$ (Section \ref{sec_eigengap_bound_L_Lrw}), is rather small in this example and decreases as the degree extreme difference grows. The edges of the normalised eigengap bounds are not attained in Fig.~\ref{fig_eigengap_bounds}. However, as discussed in Section \ref{sec_visualising_data_karate}, this is not informative on the tightness of the bound, which is valid for {\it all elements}\/ in the class of graphs $\C{1}{k}$; it is rather concerned with the particular structure of our examples. 
A feature of this comparison is the changing location of a notable eigengap, which is labelled as the $(k+9)^{\mathrm{th}}$ normalised eigengap present in the normalised Laplacian spectrum corresponding to graph $C(k)$. When clustering according to it in Section \ref{sec_clustering} it will become apparent why the location of the gap is a function of $k$. From Fig.~\ref{fig_eigengap_bounds}(l), as the degree extreme difference grows we find the difference of the $(k+9)^{\mathrm{th}}$ eigengaps of the two Laplacians to grow. In all of our examples, the $(k+9)^{\mathrm{th}}$ eigengap in the normalised Laplacian spectrum never overtakes the $10^{\mathrm{th}}$ eigengap, the only other notable eigengap. We would therefore recover the 10 connected components ahead of any other structure in the graph from the normalised Laplacian spectrum if we were to cluster according to the eigengaps in order of descending magnitude. For $A$ and $L$ this is only the case for small degree extreme differences as the $10^{\mathrm{th}}$ eigengap gets overtaken at $\dmax - \dmin = 3$ by the growing first eigengap in the adjacency spectrum and the growing $19^{\mathrm{th}}$ eigengap in the unnormalised Laplacian spectrum as seen in Fig.~\ref{fig_eigengap_bounds}(k).

The disagreement of eigengap spectra highlights that the different representation matrices recover different structures of the graphs.
The large first eigengap in the adjacency spectrum is understood as a measure of the connectedness of the graph and therefore, is seen to measure the increasing connectedness in the graph as the size of the fully connected component grows; nothing about clustering is necessarily indicated by the presence of the large first eigengap. 
The clustering structure suggested by the large $19^{\mathrm{th}}$ eigengap in the unnormalised Laplacian spectrum and the large $27^{\mathrm{th}}$ eigengap in the  normalised Laplacian spectrum of graph $C(18)$ will be discussed in Section \ref{sec_clustering}.

\begin{remark}
A further area of application for the bounds derived in this paper arises from a big data context. On large dataset the calculation of all three representation matrix spectra will come at a significant cost. However, calculating the bounds on normalised eigengap difference is inexpensive as it only requires calculation of the degree extremes, which are likely to be required at several stages of the graph's analysis. Using our inexpensive bounds practitioners will be able to anticipate the maximally different results they could have obtained using a different representation matrix. In Fig.~\ref{fig_eigengap_bounds}(b) for example the $10^{\mathrm{th}}$ eigengap of the matrices is clearly bounded away from 0 and knowledge of any one of the two spectra would suffice to say with certainty that the other spectrum will also contain a significantly larger $10^{\mathrm{th}}$ eigengap than other eigengaps in the spectrum. \hfill$\square$
\end{remark}

\begin{remark}
The examples of Fig.~\ref{fig_eigengap_bounds} show that  the eigengap spectra of the three representation matrices can deviate significantly for several eigengaps, while approximately agreeing for the majority. We find these differences to increase with growing degree extreme differences as suggested by the functional form of our bounds. Since the large eigengaps in particular are commonly used to inform  graphical analysis, a few large deviations in eigengap spectra can lead to significantly different inference as we will illustrate in Section \ref{sec_clustering}. \hfill$\square$
\end{remark}

\section{Application to Spectral Clustering}\label{sec_clustering}

 We first cluster graph $C(18)$ according to the two large eigengaps which arose from the analysis in Section \ref{sec_eigengap_bound_examples}. Then we discuss a clustering of the karate data set according to all three representation matrices. We find the choice of representation matrix to have a significant impact on the clustering outcome. 

\subsection{Spectral clustering of graph $C(18)$}

\begin{figure}[t]
\begin{center}
\includegraphics[scale=0.59,clip]{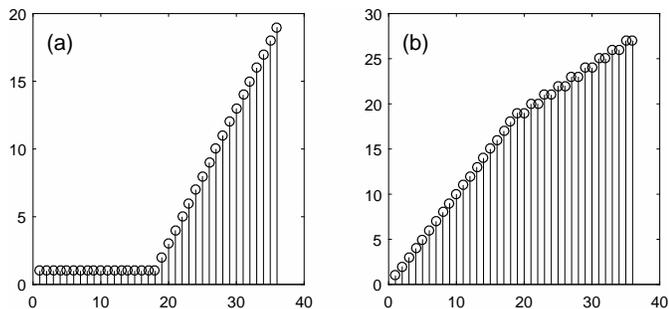}
\caption{(a) and (b) display the spectral clustering of graph $C(18)$ according to the first 19 eigenvectors of the unnormalised Laplacian $L$ and the first 27 eigenvectors of the normalised Laplacian, respectively. The x-axes display a numbering of the 36 nodes in the graph; numbers 1 to 18 correspond to nodes in the $18$-complete component and nodes 19-36 make up the 9 2-complete components. The y-axes display cluster indices ranging from 1 to 19 in (a) and 1 to 27 in (b).}
\label{fig_graphC_clustering_stem}
\end{center}
\end{figure}

We found that all representation matrices recover the 10 connected components perfectly when applying the spectral clustering algorithm to their respective first 10 eigenvectors. Additionally to the large tenth eigengap, we observed a large $19^{\mathrm{th}}$ eigengap in the \textit{unnormalised} and a large $27^{\mathrm{th}}$ eigengap in the \textit{normalised} Laplacian spectrum for graph $C(18)$ in Section \ref{sec_eigengap_bound_examples}. At first sight it is unclear what structure is suggested by these eigengaps and hence we display the clustering according to the first 19 and 27 eigenvectors of the unnormalised and normalised Laplacians in Figs.~\ref{fig_graphC_clustering_stem}(a) and (b), respectively.

From Fig.~\ref{fig_graphC_clustering_stem}(a) we see that the large $19^{\mathrm{th}}$ eigengap in the unnormalised Laplacian eigengap spectrum suggests a clustering where the 18-complete component is recovered as a single cluster and the 9 2-complete components are treated as 18 individual clusters.  
From Fig.~\ref{fig_graphC_clustering_stem}(b) we find that when clustering according to the first 27 eigenvectors of the normalised Laplacian we recover the 18-complete component as 18 clusters of one node each and each of the 2-complete components is exactly recovered.  When varying $k\in \{1, \ldots, 18\}$, the number of clusters in this clustering depends on the size of the $k$-complete component since each node in it is taken to be a separate cluster. This clarifies why we observed the large $(k+9)^{\mathrm{th}}$ eigengap in the normalised Laplacian spectrum as we varied $k$ in Fig.~\ref{fig_eigengap_bounds}. 

Essentially, the two clusterings presented in Fig.~\ref{fig_graphC_clustering_stem} are complementary, in each case clustering all nodes of the same degree in one cluster and splitting the nodes of the remaining degree into separate clusters of single nodes.

The eigengaps, in conjunction with the spectral clustering algorithm, have revealed splits of the graphs, some of which might not have been anticipated.

\subsection{Spectral clustering of the karate data set graph}

To further illustrate the impact of the representation matrix choice, we display the spectral clustering according to the first two eigenvectors of each of the three representation matrices corresponding to the karate data set. 

\begin{figure}[t]
\begin{center}
\includegraphics[scale=0.59,clip]{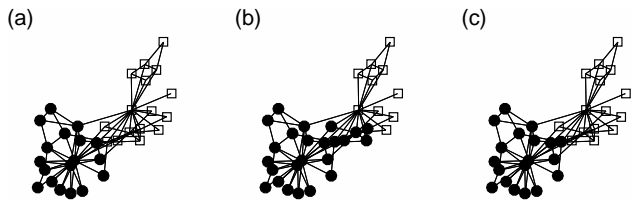}
\caption{Spectral clustering of the karate graph according to the first two eigenvectors of the adjacency matrix $A$ in (a), the unnormalised Laplacian $L$ in (b) and the normalised Laplacian $\lrw$ in (c).}
\label{fig_karate_clustering}
\end{center}
\end{figure}

Since for the karate dataset, we have a reasonably large degree extreme difference, $\dmax -\dmin =17$, we expect to see deviating results in the spectral clustering according to the different representation matrices. 

At first sight, all clusterings displayed in Fig.~\ref{fig_karate_clustering} seem sensible. The clustering according to the two normalised Laplacians agrees and hence only one of the two is displayed. The clustering according to the adjacency matrix extends the cluster marked with the unfilled, square nodes by one node, karate club member 3, in comparison to the clustering according to the normalised Laplacian. In contrast, the unnormalised Laplacian detects 5 nodes less (karate club members 2, 4, 8, 14 and 20) in the cluster marked by the unfilled square nodes, than does the normalised Laplacian. 

We find the clustering according to the adjacency matrix $A$ to agree with the true clustering suggested by the eventual split of the karate club as recorded by \cite{Zachary77}. 
The normalised Laplacians misplace one out of 34 nodes, which is known to be difficult to cluster correctly in the literature \cite{Fortunato10}. The unnormalised Laplacian however, misplaces 6 nodes, only one of which is known to us to be commonly misclustered. Hence, the unnormalised Laplacian clustering is clearly outperformed by the other two representation matrices when using the first two eigenvectors to find two communities in the karate data set. In \cite{vonLuxburg2008} the conditions under which spectral clustering using the unnormalised Laplacian converges are shown to be more restrictive than the conditions under which spectral clustering according to the normalised Laplacian converges. \cite{vonLuxburg2008} hence advocate using the normalised Laplacian for spectral clustering over the unnormalised Laplacian. Our clustering results on graph $C(18)$ and the karate data set agree with this recommendation.

\begin{remark}
For our examples, the choice of representation matrix clearly impacts the results of cluster inference via spectral clustering. We suggest considering the degree extreme difference as a parameter in graphical signal processing to infer the potential impact of the choice of representation matrix. \hfill$\square$
\end{remark}

\section{Summary and Conclusions}\label{sec:summary}

We have compared the spectra of the three graph representation matrices: the adjacency matrix $A$, the unnormalised graph Laplacian $L$ and the normalised graph Laplacian $\lrw$ and found differences in the spectra corresponding to general graphs. For all three pairs of representation matrices the degree extreme difference, $\dmax- \dmin$, was found to linearly upper bound the error when approximating eigenvalue spectra of one by an affine transformation of the spectrum of the other matrix. By considering the affine transformations we established correspondence of the largest adjacency eigenvalues to the smallest Laplacian eigenvalues. 
 We explained the monotonicity found in the bounds by partitioning the class of graphs according to their degree extremes and considering the addition/deletion of connected components to/from the graph. The bounds were visualised on several graphs, including the literature standard Zachary's karate dataset, and the first bound $e(A,L)$ was found to be tight in two out of four examples. 
 Finally, the bounds were extended to bound normalised eigengaps of the representation matrices. In examples with varying degree extremes, we saw that if the degree extreme difference is large, different choices of representation matrix may give rise to disparate inference drawn from graph signal processing algorithms; smaller degree extreme differences will result in consistent inference, whatever the choice of representation matrix. The significant differences in inference drawn from signal processing algorithms were visualised on the spectral clustering algorithm applied to model graphs and Zachary's karate graph.
As a result of this work, we hope to have increased awareness about the importance of the choice of representation matrix for graph signal processing applications.

\appendix{}

\subsection{Polynomial relation of representation matrix spectra} \label{app_poly_relation}

Here we will show that most matrix spectra can theoretically be exactly related via polynomial transforms and highlight practical issues with this approach. 

The issue of mapping one set of representation matrix eigenvalues to another can be seen as a polynomial interpolation problem on the two sets of eigenvalues. \cite[p.~390]{Horn1991} state that as long as there exists a function which maps one set of input points to the set of output points, the interpolation problem always has a unique solution and they also provide formulas to find this solution. Therefore, we find that as long as equal eigenvalues in the domain are not required to be mapped to unequal eigenvalues in the range, there exists a unique polynomial which maps one representation matrix spectrum to another.

In practice, the operation of finding the interpolating polynomial can be numerically unstable and therefore while the polynomial mapping from one spectrum to another theoretically exists, it cannot be readily obtained. For example, for the graph of the real karate data set the interpolating polynomial maps eigenvalues $\mu_{13}, \ldots, \mu_{22}$, which have a pairwise difference smaller than $10^{-14}$ onto eigenvalues of the unnormalised Laplacian which span an interval of width greater than 2. Obviously, finding the interpolating polynomial cannot be done without significant numerical error. 

Furthermore, is it unclear how one would use the polynomial map between spectra for graphical inference. Relating the polynomial coefficients to vertex degrees could lead to great insight, but is hard to achieve. Finally, we are unaware of any results which would allow calculation of the transformation without calculating both representation spectra first, while our proposed affine transformation parameters only rely on the degree extremes which are readily available from the graph.
Hence, while in theory, in some cases, there exists a more precise relation of the representation matrix spectra, our simple affine transformations have significant advantages.

\subsection{Weyl's inequality and the bound of (\ref{eqn_L_error_bound})}\label{app_weyl_bound}
In (\ref{eqn_defn_L_hat}) of Section~\ref{sec_bd_A_L} we define $\hat{L} = L +d_1 I - D$. Denote the error between $\hat{L}$ and $L$ by  $\Delta  = d_1I-D,$ and the error eigenvalues as follows: $\lambda_1(\Delta) \leq \lambda_2(\Delta) \leq ... \leq \lambda_n(\Delta)$.
Weyl's inequality states that, for $i \in \left\{1,2,\ldots,n\right\},$
$$
\lambda_i + \lambda_1\left( \Delta\right)  \leq \lambda_i(\hat{L}) \leq \lambda_i + \lambda_n\left( \Delta\right) . 
$$
So, in particular,
\begin{flalign*}
\lambda_1\left( \Delta\right) & \leq \lambda_i(\hat{L}) - \lambda_i\leq  \lambda_n\left( \Delta\right).\\
\Rightarrow d_1 - d_{\max} & \leq \lambda_i(\hat{L}) - \lambda_i \leq d_1 - d_{\min}.
\end{flalign*}
Now, set $d_1=(d_{\max}+ d_{\min})/{2}$ to obtain the result derived in  (\ref{eqn_L_error_bound}) of 
Section~\ref{sec_bd_A_L}, which is a bound on the absolute difference.

This result extends the application of Weyl's inequality to the two representation matrix spectra, as is done in \cite[p.~71]{vanMieghem2011}, by transforming one of the two spectra.

\subsection{Choosing $c_2$}\label{app:choosec2}
In what follows we refer to the  terms $|c_2| (d_{\max} - d_{\min})/{2}$ and $2 \max\left( \left\lvert d_{\max} c_2 - 1 \right\rvert, \left\lvert d_{\min} c_2 -1 \right\rvert \right)$ as term 1 and term 2, respectively.
We proceed by analysing the bound in the three different intervals 
$$
\Big(-\infty,0 \Big),\left\lbrack0,\frac{2}{d_{\max} + d_{\min}}\right\rbrack,\left(\frac{2}{d_{\max} + d_{\min}}, \infty\right).
$$
For $c_2 \in \Big(-\infty,0 \Big)$ we have, 
\begin{equation*}
\max\left( \left\lvert d_{\max} c_2 - 1 \right\rvert, \left\lvert d_{\min} c_2 -1 \right\rvert \right) = -d_{\max} c_2 +1,
\end{equation*}
and both terms 1 and 2 decrease as  $c_2$ increases.
 Hence, the overall bound is decreasing and thus, the value of $c_2$ minimising the bound, satisfies $0\leq c_2$.

For $c_2 \in \left({2}/(d_{\max} + d_{\min}), \infty\right)$ we have 
\begin{equation*}
\max\left( \left\lvert d_{\max} c_2 - 1 \right\rvert, \left\lvert d_{\min} c_2 -1 \right\rvert \right) = d_{\max} c_2 -1.
\end{equation*}
Hence, both terms 1 and 2 are increasing, and the value of $c_2$ minimising the bound satisfies $c_2\leq {2}/(d_{\max} + d_{\min}).$

Finally, for $ c_2 \in \left\lbrack 0,{2}/(d_{\max} + d_{\min})\right\rbrack,$ we have 
\begin{equation*}
\max\left( \left\lvert d_{\max} c_2 - 1 \right\rvert, \left\lvert d_{\min} c_2 -1 \right\rvert \right) = -d_{\min} c_2 + 1.
\end{equation*}

Hence, term 2 is decreasing while term 1 is increasing. We will now determine which of the two terms defines the monotonicity of the bound by comparing slopes.
\begin{flalign*}
\text{Term 1 dominates term 2}
&\iff \frac{d_{\max} - d_{\min}}{2} - 2 d_{\min}> 0\\
&\iff d_{\max} > 5 d_{\min}.
\end{flalign*}
Hence, if $d_{\max} > 5 d_{\min}$, the bound is increasing in the interval $\left\lbrack 0,{2}/(d_{\max} + d_{\min})\right\rbrack$. Therefore, $c_2=0$ minimises the bound, so that 
\begin{align*} 
&\left\lvert \eta_i(\tilde{L}_{rw}) - \eta_i \right\rvert
 \leq \Big(|c_2| \frac{d_{\max} - d_{\min}}{2}  \\ 
&+  2 \max\left( \left\lvert d_{\max} c_2 - 1 \right\rvert, \left\lvert d_{\min} c_2 -1 \right\rvert \right) \Big)_{c_2=0}\\ 
&=2 \,\, \text{ for }d_{\max} > 5 d_{\min}. \numberthis \label{eqn_A_Lrw_bd1}
\end{align*}
Otherwise, if  $d_{\max} \leq 5 d_{\min}$, we have a decreasing bound. This results in a minimal bound at $c_2={2}/(d_{\max} + d_{\min})$ and the bound value
\begin{align*} 
&\left\lvert \eta_i(\tilde{L}_{rw}) - \eta_i \right\rvert <  \Big(|c_2| \frac{d_{\max} - d_{\min}}{2}  \\
& + 2 \max\left( \left\lvert d_{\max} c_2 - 1 \right\rvert, \left\lvert d_{\min} c_2 -1 \right\rvert \right) \Big)_{c_2=\frac{2}{d_{\max} + d_{\min}}}\\  
&= \frac{d_{\max} - d_{\min}}{d_{\max} + d_{\min}} + 2 \frac{d_{\max} - d_{\min}}{d_{\max} + d_{\min}} \numberthis \label{eqn_A_Lrw_optim_as_before}\\
&=3 ~\frac{d_{\max} - d_{\min}}{d_{\max} + d_{\min}} \leq2 \,\, \text{ for }d_{\max} \leq 5 d_{\min}. \numberthis \label{eqn_A_Lrw_bd2}
\end{align*}
The result in \eqnref{eqn_A_Lrw_optim_as_before} follows from plugging $c_2$ into term 1 and recognising that term 2 is degenerate since both its arguments are equal. We denote the resulting bound by $e'\left( A, \lrw \right):$
\begin{equation}\label{eq:edash}
e'\left( A, \lrw \right) \EqualDef
\begin{cases}
 3 ~\frac{d_{\max} - d_{\min}}{d_{\max} + d_{\min}}, &\text{for } d_{\max} < 5 d_{\min};\\
2, & \text{otherwise.}
\end{cases}
\end{equation}
For $d$-regular graphs, for which  $d_{\max}=d_{\min}=d,$ the upper result applies, and we see that the bound is again zero. For a graph with disconnected nodes $d_{\min}=0,$ the lower result applies, and the bound is 2; from (\ref{eqn_A_Lrw_bd2}) this is the maximal bound and unlike when comparing $L$ and $\lrw,$ this bound is already reached for $d_{\max} \geq 5 d_{\min},$ and not in the limit $d_{\max} \rightarrow \infty$ for fixed $d_{\min}.$

From (\ref{roleofc2}) we see that if $c_2=0$, then the transformation is degenerate, non-invertible and all information is lost in the process. Hence we will reject the bound value $e'\left( A, \lrw \right) = 2 $ for $d_{\max} > 5 d_{\min}$ even though it results in the minimal bound value -- the lowest bound value achieved under our analysis framework and not the absolute minimal bound, i.e., a tight bound.  So, we adopt the bound given in (\ref{eq:eord}), namely,
\begin{align*}
e\left( A, \lrw \right) &\EqualDef  3 ~\frac{d_{\max} - d_{\min}}{d_{\max} + d_{\min}} .
\end{align*}

\section*{Acknowledgment}

The work of Johannes Lutzeyer is supported by the EPSRC (UK).

\end{document}